\documentclass[a4paper,12pt]{article}
\usepackage{amsfonts,amssymb}
\usepackage{latexsym}
\usepackage{mathrsfs}
\usepackage[colorlinks=true,linkcolor=red]{hyperref}

\oddsidemargin
\topmargin
  \usepackage{amsthm}
  \usepackage{amsmath}

\usepackage{amssymb}

  \newtheorem{theorem}{Theorem}[section]
  \newtheorem{corollary}[theorem]{Corollary}
  \newtheorem{proposition}[theorem]{Proposition}
  \newtheorem{lemma}[theorem]{Lemma}

  \theoremstyle{definition}
  \newtheorem{definition}[theorem]{Definition}
  \newtheorem{remark}[theorem]{Remark}

  \numberwithin{equation}{section}

  \title{\sc Who Says What: Strategic Communication in Networks \\[2ex] 
  }
 
\author{
{\bf Anuj Bhowmik}\thanks{Indian Statistical Institute, 203 B.T. Road, Kolkata 711108, India. Email: anujbhowmik09@gmail.com}
\and
{\bf Saptarshi P. Ghosh}\thanks{Indian Institute of Technology, Bombay, Maharashtra, India. Email: spghosh@iitb.ac.in}
\and
{\bf Iman Kundu}\thanks{Indian Statistical Institute, 203 B.T. Road, Kolkata 711108, India. Email: imankundu98161@gmail.com}
\and
{\bf Sandipan Saha}\thanks{Indira Gandhi Institute of Development and Research, Maharashtra, India. Email: sandipan.saha.10@gmail.com}
}

\date{}

\begin{document}

  \maketitle
  

\begin{abstract}
  
We study a multi-player communication-coordnation game in which privately informed agents communicate through a social network and choose their optimal actions while balancing it against their own bliss points and that of the neighbours actions. We show that equilibrium actions are convex combinations of the expected bliss points of all players, with weights being determined by a degree-discounted Bonacich centrality measure. We also characterize when players report truthfully, over-report, or under-report their private information, and show that the incentive to misreport depends critically on the receiver’s position in the network. A striking implication is that even under identical preferences fully revealing communication may not realise and information transmission may remain coarse in a star network while becoming almost fully informative in a complete network. Finally, when biases are endogenously chosen, strategic responses to neighbours’ biases are governed by their centralities, and unbiasedness may emerge as a Nash equilibrium.
\end{abstract}


\medskip


\section{Introduction}\label{Sec:1}

People often share information with those whose actions they
must coordinate. Consider two interns who work on the same
project but report to different managers. Each has information
relevant to the project, but their responsibilities may lead
them to prefer different decisions. Whether they share
information truthfully depends on both these differences and
their communication links: whom they can contact and which
messages they observe. This paper examines how the network
shapes reporting incentives, showing that truthful communication
depends on both preferences and the structure of interactions.

Our study is somewhat related to the literature on strategic information
transmission. Crawford and Sobel (1982) show how preference
disparity between an informed sender and a receiver  limits communication between them, with information transmitted being linked to partitions of
the sender's type space. In settings with several decision
makers, communication also depends on the pattern
of interactions embedded among the decision makers. Hagenbach and Koessler (2010) study strategic
communication before a coordination game and characterise
networks that are formed on basis of truthful information transmission. Galeotti,
Ghiglino, and Squintani (2013) show that a player's ability to
communicate truthfully depends on both preferences and the
information received by other decision makers. These studies
reflects that communication strategies are not merely influenced through bilateral preference differences alone but also are influenced by the surrounding pattern of interaction.

We build on this insight by analysing communication among a set of decision makers connected through
an exogenously given network that governs their access to information
and drives their incentives to coordinate. in contrast to the existing literature, rather than treating a link between decision makers as
truthful information transmission, we treat it as a medium
to communicate and examine how players choose to report their private information.
Our focus is on how the given network influences continuation actions,
communication in form of nformation transmission, and the direction and magnitude
of misreporting. Each player acts as both a sender and a receiver,
while also choosing her optimal decision.

The paper relates to the literature connecting influence of network
on equilibrium behaviour. Ballester, Calv\'o-Armengol,
and Zenou (2006) establish a relationship between equilibrium
actions and Bonacich centrality in games with local
complementarities. De Mart\'i and Zenou (2015) examine network
games with incomplete information, where players must form
beliefs about others' information. In our setting, these beliefs
are shaped by strategic communication choices. Consequently, how
one decision maker influences another depends on both the network
and the information conveyed in the form of messages.

We device a two-stage game in which each player receives
private information about a state of the world, which equalsthe sum of all
players' private signals. Players first exchange messages with their neighbours in the network and
then choose actions. Each player wants her action to be close
to the state of the world shifted in the direction of her personal bias, while also attempting to align 
with her neighbours' actions. In the baseline model,
communication is local and private i.e., players communicate only
with neighbours and do not observe messages exchanged between
other pairs. Thus, a player must form beliefs about the
information available to others and hence the state of the world. We also examine special instances where 
public communication is observed and a setting in which a neighbour takes
reports at face value.

Our first result characterises equilibrium actions following
communication. For a fixed communication strategy profile, the
continuation game has a unique square-integrable equilibrium
action profile, up to almost-sure equality. Its representation
uses an operator-valued centrality matrix that combines network
interactions with players' conditional expectations. This
representation captures how direct and indirect
communication channels influence information and beliefs about
others' information held by a player in equilibrium. It also highlights the inability of numerical centrality weights alone
to describe equilibrium actions under
incomplete information.

We next examine the communication stage with Bayesian recipients. For
a complete network with simultaneous public communication,
independent uniform types, and a common unit coordination
weight, we derive the partition cutoff equation and the maximum
possible number of communication intervals of a players' type space. Reporting
incentives depend on a player's aggregate bias difference from
the other players. As all biases approach towards a common value,
the network allows equilibrium partitions with arbitrarily
many intervals of the type space for every player.

The three-player star illustrates the additional role of
information access. The centre communicates a common message
to both the peripheral players, but each such peripheral player observes neither the other player's
report nor its private information. The resulting cutoff
equation of the peripheral players contain a type-dependent distortion that is absent
from the complete-network benchmark. This difference shows
how the network architecture changes reporting incentives through the
information players possess and the actions they anticipate,
even after accounting for their biases. The star analysis
distinguishes the peripheral players' optimal reporting conditions from the
centre's conditional comparison when incoming reports are
held fixed.

For communication with a naive neighbour, we introduce an
absolute cost of misreporting and characterise communication through means of truthful
reporting, over-reporting, and under-reporting. We show that truthfulness is
optimal when the difference between the sender's action under
a truthful report and her expectation of the recipient's
action is sufficiently small when other
messages and continuation strategies are held fixed, 
The threshold depends on the
cost of deception, the strength of coordination, and the
recipient's response to the report. Network structure therefore
influences both the difference in expected action and the
effectiveness of a message in changing the recipient's action by influencing his or her beliefs.
We also derive the optimal report and establish its monotonicity
in the sender's type under the stated assumptions, allowing for
pooling at the boundaries of the message space.

Finally, we allow players to choose their biases. Holding the
information structure and other players' biases fixed, we
characterise how a player's optimal bias responds to a change
in another player's bias. The response reflects the latter's
influence on the former's neighbours, linking bias choice to
network position. Together, the results identify the channels
through which preferences, information access, and coordination
jointly determine communication: bias differences matter, but
their consequences depend on the network in which players
interact.

The rest of the paper is organized as follows. Section \ref{Sec:2} discusses our framework and talks about the game. Section \ref{Sec:3} introduces the important notion of centrality and discusses the results regarding centrality. Section \ref{Sec:4} introduces the notion of equilibrium in our game and highlights our findings about the first-stage and second-stage equilibrium characterisation. Section \ref{Sec:5} talks about endogenous biases in our setup. Section \ref{Sec:6} concludes.

\section{The Model}\label{Sec:2}

Let $N=\{1,2,\cdots,n\}$ denote the set of players. Players are connected through a social network represented by a graph $g$, where $g_{ij}$ denotes the link between players $i$ and $j$. For any $i,j\in N$, let $g_{ij}=1$ if $i$ and $j$ are connected and $g_{ij}=0$ otherwise. More generally, one may allow $0\leq g_{ij}\leq 1$, in which case $g_{ij}$ represents the strength of the link between players $i$ and $j$. By convention, $g_{ii}=0$ for all $i\in N$. The set of neighbours of player $i$ is denoted by $N_i:=\{j\in N:g_{ij}=1\}$, and $n_i:=|N_i|$ denotes the number of her neighbours or degree of player $i$. Let $G:=[g_{ij}]_{i,j=1}^{n}$ denote the adjacency matrix of the network. Throughout, we assume that the network is undirected, so that $g_{ij}=g_{ji}$ for all $i,j\in N$.

Each player $i$ chooses an action $a_i\in A_i=\mathbb R$. Let $A:=\prod_{i=1}^{n}A_i$ denote the set of action profiles, and let $a=(a_1,\cdots,a_n)\in A$ denote a generic action profile. Each player's payoff depends on the action profile, the state of the world $\theta$, and her communication strategy. There exists a state of the world  which is unknown to all players. Each player $i\in N$ receives private partial information $k_i\in K_i$ about $\theta\in \Theta:=[0, n]$, where $K_i$ is a random variable on $[0,1]$ with distribution function $F_i(\cdot)$ and density function $f_i(\cdot)$. Let $K:=\prod_{i=1}^{n}K_i$ denote the space of all possible profiles of private information. It is common knowledge that each player's private information lies in $[0,1]$, and the distribution functions $F_1,\cdots,F_n$ are also common knowledge. Denote the profile of private information by $\textbf{k}:=(k_i)_{i\in N}$ and define the state of the world as $\theta(\textbf{k}):=\sum_{i\in N}k_i$. Following Hagenbach and Koessler (2009), we adopt this additive specification to abstract from inessential complications and focus on the strategic features of the communication game.

\subsection{Communication Game}

After observing their private information and before choosing actions, players exchange costless, private messages through network links. Player $i$ sends a message $m_{ij}\in M_{ij}$ to each neighbour $j\in N_i$, where $M_{ij}\subseteq[0,1]$ denotes the set of messages she can send to player $j$. Communication is restricted to neighbours, and each message is observed only by its sender and receiver.

Let $M_i:=\prod_{j\in N_i}M_{ij}$ denote the set of message profiles that player $i$ can send to her neighbours, and let $\textbf{m}_i:=(m_{ij})_{j\in N_i}$ denote a particular profile of messages sent by player $i$ to her neighbours. Similarly, let $M_i^r:=\prod_{j\in N_i}M_{ji}$ denote the set of message profiles that player $i$ can receive from her neighbours, and let ${\bf m}_i^r:=(m_{ji})_{j\in N_i}$ denote a particular profile of messages received by player $i$ from her neighbours. Thus, each player observes the messages she sends and receives, but not the messages exchanged between other pairs of players.

\subsection{Coordination Game}

Following the communication stage, each player chooses an action in the second stage, while aligning it to the state of the wolrd shifted in the direction of her bias and that of her neighbours actions. The utility of player $i$ is given by $u_{i}: \Theta \times \prod\limits_{j \in N_{i} \cup \{i\}}A_{j} \to \mathbb{R}$, where
\begin{equation}\label{eqn:utility1}
u_{i}(\theta, (a_{j})_{j \in N_{i} \cup \{i\}})= -(a_{i} - \theta - b_{i})^{2} -\sigma_i\sum\limits_{j \in N_{i}}(a_{i} - a_{j})^{2},
\end{equation}
where $\sigma_{i}>0$ for all $i\in N$ and $b_{i}\in\mathbb{R}$ denotes the bias of player $i$. The first term captures the loss from choosing an action that differs from player $i$'s bliss point, $\theta+b_i$, which is determined by the state of the world and her individual bias. The second term captures the motive for coordination with her neighbours, in which player $i$ incurs a loss whenever her action differs from that of a neighbour, with $\sigma_i$ measuring the strength of this coordination. Thus, a player's choice of action reflects a trade-off between remaining close to her own bliss point and coordinating with the actions of her neighbours.


\medskip
\noindent
Finally, the framework can be extended to weighted networks by allowing $0\leq g_{ij}\leq 1$, where $g_{ij}$ measures the strength of the connection between players $i$ and $j$. In such a formulation, communication takes place whenever $g_{ij}>0$, while the influence of neighbour $j$ on player $i$ depends on the strength of their link. This can be incorporated into the coordination term by attaching a weight of $g_{ij}$ to the loss from difference in actions. 

\section{The Notion of Centrality}\label{Sec:3}

We introduce the centrality measures used in our analysis. Let $G$
denote the adjacency matrix of the undirected network $g$. For
$w\geq0$ satisfying $w\rho(G)<1$, where $\rho(G)$ denotes the
spectral radius of $G$, define
\[
B(w,G):=[I-wG]^{-1}
=\sum_{k=0}^{\infty}w^kG^k.
\]
The parameter $w$ acts as a decay factor, assigning progressively lower weights to longer walks in the network.
The $(i,j)$-th entry, $b_{ij}(g,w)=\sum_{k=1}^{\infty}w^k g_{ij}^{[k]}$, measures the weighted number of walks 
from $i$ to $j$, where walks of length $k$ receive weight $w^k$.

\begin{definition}
For a network $g$ with adjacency matrix $G$ and a parameter
$w\geq0$ satisfying $w\rho(G)<1$, the vector of
\emph{Bonacich centralities} is
\[
{\bf b}(g,w):=[I-wG]^{-1}{\bf 1},
\]
\end{definition}
 where ${\bf 1}$ denote the $n$-dimensional vector of ones. The Bonacich centrality of node $i$ is given by $b_i({\bf g},w)=\sum_{j\in N}b_{ij}({\bf g},a)$ and measures the total weighted number of walks originating from $i$. 

We next introduce a degree-adjusted version of the network, which is a normalization that incorporates players'
degrees and coordination parameters. To this end, define $L=[l_{ij}]_{i,j=1}^{n}$ by
\[
l_{ij}:=
\begin{cases}
\displaystyle\frac{1}{1+\sigma_i n_i}, & \text{if }i=j,\\[6pt]
0, & \text{if }i\neq j.
\end{cases}
\]
Define the normalized network matrix by
\[
G^N
:=
\left[
\frac{\sigma_i g_{ij}}{1+\sigma_i n_i}
\right]_{i,j=1}^{n}
=
L\operatorname{diag}(\sigma_1,\cdots,\sigma_n)G.
\]
A similar normalization appears in Ghiglino \textit{et al.} (2010). Under this transformation, the weight that player $i$ assigns to the link with player $j$ is $\frac{\sigma_ig_{ij}}{1+\sigma_i n_i}$. Thus, for fixed $\sigma_i$, as the number of neighbours of player $i$ increases, the weight assigned to each individual connection decreases. Moreover,
\[
\rho(G^N)
\leq\|G^N\|_\infty
=
\max_{i\in N}\frac{\sigma_i n_i}{1+\sigma_i n_i}
<1.
\]
Consequently, the matrix
\begin{equation}\label{eqn:centrality}
D(1,G^N):=(I-G^N)^{-1}
=\sum_{k=0}^{\infty}(G^N)^k
=[d_{ij}]_{i,j=1}^{n}
\end{equation}
is well defined and nonnegative. Under the convention that
$(G^N)_{ij}$ measures the influence of $j$ on $i$, the entry
$d_{ij}$ measures the number of weighted walks from player $j$ to player $i$ and thus captures the influence of $j$ on $i$ in the Bonacich-centrality sense.

Although $G$ is symmetric, $G^N$ need not be symmetric.
Symmetry requires
$\sigma_i/(1+\sigma_i n_i)=\sigma_j/(1+\sigma_j n_j)$
whenever $g_{ij}=1$. In particular, this condition holds
when the network is regular\footnote{A graph is regular if $n_i=k$ for every $i\in N$ and for some $k\in\{0,\cdots,n-1\}$.} 
 and all players have the same
coordination parameter. 

We normalize $d_{ij}$ by the cardinality of player $j$'s augmented neighbourhood,\footnote{Player $j$'s augmented neighbourhood consists of player $j$ herself and her neighbours and therefore has cardinality $1+n_j$.} defining $c_{ij}=\sigma_jd_{ij}/(1+\sigma_jn_j)$. In matrix form,
\begin{equation}\label{eqn:centralityprime}
C(1,G^N):=D(1,G^N)L=[c_{ij}]_{i,j=1}^{n}.
\end{equation}
This is the matrix of equilibrium action weights under complete
information, where the weight that the player $i$ places on the
bliss point of player $j$ is the $(i,j)^{{\rm th}}$-entry of the matrix $C(1, G^N)$ (refer to Theorem \ref{thm:equilibriumaction}).
As the following lemma shows the sum of each row is equals to one, so each player's action is a
convex combination of the bliss points of all the players. With heterogeneous
coordination parameters, its entries satisfy a weighted symmetry
identity; ordinary symmetry obtains when the coordination
parameter is common.

\newpage

\begin{lemma}
Let $G^N$ and $C(1,G^N)$ be defined as above. Then:
\begin{enumerate}
    \item[\emph{(i)}]
    $C(1,G^N)$ is nonnegative and
    $\sigma_jc_{ij}=\sigma_ic_{ji}$ for all $i,j\in N$;
    \item[\emph{(ii)}]
    $c_{jj}>c_{ij}$ for all distinct $i,j\in N$;
    \item[\emph{(iii)}]
    $\sum_{j\in N}c_{ij}=1$ for every $i\in N$.
\end{enumerate}
If $\sigma_i=\sigma>0$ for every $i\in N$, then
$C(1,G^N)$ is symmetric,
$c_{ii}>c_{ij}=c_{ji}$ for all distinct $i,j\in N$, and
\[
0<c_{ii}\leq\frac{1+\sigma}{1+\sigma n_i}.
\]
Consequently, along any sequence of networks with a common,
fixed coordination parameter $\sigma$, $c_{ii}\to0$
whenever $n_i\to\infty$.\footnote{Note that $c_{ii}$ depends on $\sigma$ and $n_i$ as follows from the definition 
of $C(1, G^N)$.}
\end{lemma}

\begin{proof}
(i) Nonnegativity follows from
$C(1,G^N)=\sum_{k=0}^{\infty}(G^N)^kL$.
The same expansion gives
$c_{ii}\geq1/(1+\sigma_i n_i)>0$.
Furthermore,
\[
C(1,G^N)\operatorname{diag}(\sigma_1,\ldots,\sigma_n)
=
\left[
\operatorname{diag}\left(
n_1+\frac{1}{\sigma_1},\ldots,
n_n+\frac{1}{\sigma_n}
\right)-G
\right]^{-1}.
\]
Since $G$ is symmetric, the matrix on the right is symmetric.
Comparing its $(i,j)$- and $(j,i)$-entries yields
$\sigma_jc_{ij}=\sigma_ic_{ji}$.

\medskip
\noindent
(ii) Comparing entries in $(I-G^N)C(1,G^N)=L$ gives
\[
c_{ij}=
\begin{cases}
\displaystyle
\frac{\sigma_i}{1+\sigma_i n_i}
\sum_{k\in N_i}c_{kj},
& \text{if }i\neq j;\\[8pt]
\displaystyle
\frac{1}{1+\sigma_i n_i}
+\frac{\sigma_i}{1+\sigma_i n_i}
\sum_{k\in N_i}c_{ki},
& \text{if }i=j.
\end{cases}
\]
Fix $j\in N$. If the maximum entry in column $j$ were
attained at some $i\neq j$, then $c_{ij}\geq c_{jj}>0$
and
\[
c_{ij}
=
\frac{\sigma_i}{1+\sigma_i n_i}
\sum_{k\in N_i}c_{kj}
\leq
\frac{\sigma_i n_i}{1+\sigma_i n_i}c_{ij}
<c_{ij},
\]
which is a contradiction. Therefore, $c_{jj}>c_{ij}$ for every $i\neq j$.

\medskip
\noindent
(iii)
For every $i\in N$,
\[
\bigl[(I-G^N){\bf 1}\bigr]_i
=
1-\frac{\sigma_i n_i}{1+\sigma_i n_i}
=
\frac{1}{1+\sigma_i n_i}
=
(L{\bf 1})_i.
\]
Therefore, $(I-G^N){\bf 1}=L{\bf 1}$, and
\[
C(1,G^N){\bf 1}
=(I-G^N)^{-1}L{\bf 1}
={\bf 1}.
\]
Consequently, $\sum_{j\in N}c_{ij}=1$ for every $i\in N$.

Finally, suppose that $\sigma_i=\sigma>0$ for every $i$.
Part (i) implies symmetry, which together with
part (ii) gives
$c_{ii}>c_{ij}=c_{ji}$ for every $i\neq j$.
Using the symmetry, positivity of the diagonal elements,
and part (iii), we obtain
\[
0<c_{ii}
=
\frac{1+\sigma\sum_{k\in N_i}c_{ki}}{1+\sigma n_i}
=
\frac{1+\sigma\sum_{k\in N_i}c_{ik}}{1+\sigma n_i}
\leq
\frac{1+\sigma}{1+\sigma n_i}.
\]
The right-hand side converges to zero as $n_i\to\infty$
for fixed $\sigma>0$.
\end{proof}

\section{Equilibrium}\label{Sec:4}

In this section we talk about the communication stage. Each player observes her private information and the messages she sends and receives, but not the messages exchanged between other players. So communication is private in our setup. Thus, her action therefore depends on her beliefs about other players' private information and communication elsewhere in the network. Since $\theta({\bf k})=\sum_{i\in N}k_i$, her beliefs about the state are induced by her beliefs about ${\bf k}_{-i}$.

A pure communication strategy for player $i$ is a measurable mapping $\beta_i:K_i\to M_i$, where $\beta_i(k_i)=(\beta_i^j(k_i))_{j\in N_i}$ and $\beta_i^j(k_i)$ is the message sent to neighbour $j$. An action strategy is a measurable mapping $\tau_i:K_i\times M_i\times\mathscr M_i\to A_i$, where $\tau_i(k_i,{\bf m}_i,{\bf m}_i^r)$ specifies her action after observing $k_i$, sending ${\bf m}_i$, and receiving ${\bf m}_i^r$. Write $\beta=(\beta_i)_{i\in N}$ and $\tau=(\tau_i)_{i\in N}$.

For a private information profile ${\bf k}$ and a message profile ${\bf m}=({\bf m}_i)_{i\in N}$, define the induced action of player $i$ by
\[
a_i^\tau({\bf k},{\bf m})
:=\tau_i\bigl(k_i, (m_{ij})_{j \in N_i}, (m_{ji})_{j\in N_i}\bigr).
\]
The full profiles ${\bf k}$ and ${\bf m}$ describe the outcome; player $i$'s action depends only on her own observations. To simplify notation, write her induced payoff as
\[
V_i({\bf k},{\bf m};\tau)
:=
u_i\left(\theta({\bf k}),
\bigl(a_j^\tau({\bf k},{\bf m})\bigr)_{j\in N_i\cup\{i\}}
\right).
\]

Let $\mu_i(\cdot\mid k_i,{\bf m}_i,{\bf m}_i^r)$ denote player $i$'s belief over $({\bf k}_{-i},{\bf m}_{-i})$ at her action-stage history. These beliefs assign probability one to message profiles consistent with the messages she has received. All strategies and beliefs below are assumed to yield finite expected quadratic losses.

\begin{definition}\label{defn:eqm}
A pure-strategy perfect Bayesian equilibrium is an assessment $(\beta^*,\tau^*,\mu^*)$ satisfying the following conditions.

\begin{enumerate}
\item[(i)] \textit{Optimal communication.}
For every $i\in N$ and $k_i\in K_i$,
\[
\beta_i^*(k_i)\in
\arg\max_{{\bf m}_i\in M_i}
\mathbb E\left[
V_i\left(
{\bf k},
({\bf m}_i,\beta_{-i}^*({\bf k}_{-i}));
\tau^*
\right)
\,\middle|\, k_i
\right].\footnote{Here, $\beta_{-i}^*({\bf k}_{-i})=(\beta_j^*(k_j))_{j\neq i}$, and expectations are taken under the common prior conditional on $k_i$.}
\]
Each player chooses her outgoing messages optimally, anticipating
subsequent play according to $\tau^*$.

\item[(ii)] \textit{Optimal actions.}
For every $i\in N$ and every feasible history
$(k_i,{\bf m}_i,{\bf m}_i^r)$,
\[
\tau_i^*(k_i,{\bf m}_i,{\bf m}_i^r)
\in
\arg\max_{a_i\in A_i}
\mathbb E_{\mu_i^*}\left[
u_i\left(
\theta({\bf k}),
\left(a_i,
\bigl(a_j^{\tau^*}({\bf k},{\bf m})\bigr)_{j\in N_i}
\right)
\right)
\,\middle|\,
k_i,{\bf m}_i,{\bf m}_i^r
\right].
\]
Here, the action
profile inside $u_i$ assigns action $a_i$ to player $i$ and
action $a_j^{\tau^*}({\bf k},{\bf m})$ to each neighbour $j$.

\item[(iii)] \textit{Consistent beliefs.}
For every $i\in N$ and every outgoing message profile
${\bf m}_i\in M_i$, the belief
$\mu_i^*(\cdot\mid k_i,{\bf m}_i,{\bf m}_i^r)$
is a regular conditional distribution of $\bigl({\bf k}_{-i},\beta_{-i}^*({\bf k}_{-i})\bigr)$
given $\left(k_i,(\beta_j^{*i}(k_j))_{j\in N_i}\right)=(k_i,{\bf m}_i^r)$,
under the common prior and the other players' equilibrium
communication strategies. This requirement holds almost surely under the induced distribution
of private information and received messages. At histories where
Bayesian updating does not determine beliefs, they are specified
as part of the assessment and must remain consistent with the
player's observations.
\end{enumerate}
\end{definition}

\begin{remark}
Condition (i) requires each player to choose her outgoing messages optimally, holding the other players' communication strategies fixed and anticipating subsequent play according to $\tau^*$. Condition (ii) requires her action to maximize expected utility at every feasible action-stage history. The distinction between the two stages lies in the information available: communication decisions are evaluated under the common prior conditional on $k_i$, whereas action decisions are evaluated under posterior beliefs conditional on the player's private information and communication history. Condition (iii) requires these beliefs to agree with Bayesian updating wherever it applies.
\end{remark}

\begin{remark}
Since private information is independent across players, Bayesian updating along equilibrium play operates separately for each neighbour: player $i$ updates her belief about $k_j$ using the message received from $j$, while retaining the prior distributions of non-neighbours. Beliefs about unobserved messages are then induced by the other players' communication strategies. Her own outgoing messages provide no additional information about others' private information or messages. Accordingly, the Bayesian consistency requirement applies to every outgoing message profile, including deviations from her equilibrium communication strategy.
\end{remark}

\subsection{Characterisation of Equilibrium Action Strategies}

Fix an equilibrium communication strategy profile
$\beta^*=(\beta_j^*)_{j\in N}$. We characterise the continuation
equilibrium induced by this profile. For each $i\in N$, write
\[
\widetilde{\beta}_i^*({\bf k})
:=
\left(
\beta_i^*(k_i),
(\beta_j^{*i}(k_j))_{j\in N_i}
\right)
\]
for the profiles of messages sent and received by player $i$.
Let $\mathcal F_i$ denote the $\sigma$-algebra generated by
$(k_i,\widetilde{\beta}_i^*({\bf k}))$, and let $E_i$ be the
associated conditional expectation operator under the common prior.
Therefore, for any integrable random variable $X$,
\[
E_i(X)
:=
\mathbb E\left[
X\mid k_i,\widetilde{\beta}_i^*({\bf k})
\right].
\]
In particular, player $i$'s posterior expectation of the state is
$\theta_i({\bf k}):=E_i(\theta({\bf k}))$, where
$\theta({\bf k})=\sum_{\ell\in N}k_\ell$.
Consider the Banach space
$\mathcal H:=\prod_{i\in N}L^2(\mathcal F_i)$, where square
integrability is with respect to the common prior, equipped with
the norm $\|x\|:=\max_{i\in N}\|x_i\|_2$. Recall that 
\[
L:=
\operatorname{diag}\left(
\frac{1}{1+\sigma_1n_1},\cdots,
\frac{1}{1+\sigma_nn_n}
\right)
\; \mbox{ and define }\; 
\mathcal G^N_{ij}
:=
\frac{\sigma_i g_{ij}}{1+\sigma_i n_i}E_i.
\]
Here, $L$ acts componentwise by scalar multiplication, and
$\mathcal G^N=[\mathcal G^N_{ij}]_{i,j=1}^n$ acts on
$\mathcal H$ according to
\[
(\mathcal G^N x)_i
=
\frac{\sigma_i}{1+\sigma_i n_i}
\sum_{j\in N_i}E_i(x_j).
\]
Define the operator-valued centrality matrix by
\begin{equation}\label{eqn:operatorcentrality}
\mathcal C:=(I-\mathcal G^N)^{-1}L
=[\mathcal C_{ij}]_{i,j=1}^n.
\end{equation}
The inverse exists by the contraction argument below. Each
$\mathcal C_{ij}$ is a bounded linear operator from
$L^2(\mathcal F_j)$ to $L^2(\mathcal F_i)$.

\begin{theorem}\label{thm:equilibriumaction}
Fix an equilibrium communication strategy profile $\beta^*$.
The induced continuation game admits a unique square-integrable
equilibrium action profile, up to almost-sure equality. For every
$i\in N$, it satisfies
\begin{equation}\label{eqn:equilibriumactionoperator}
\tau_i^*\left(k_i,\widetilde{\beta}_i^*({\bf k})\right)
=
\sum_{j\in N}
\mathcal C_{ij}\left(\theta_j({\bf k})+b_j\right)
\mbox{ almost surely, }
\end{equation}
where $\theta_j({\bf k})=E_j(\theta({\bf k}))$ and
$\mathcal C=(I-\mathcal G^N)^{-1}L$.
\end{theorem}

\begin{proof}
Let $x=(x_i)_{i\in N}\in\mathcal H$ denote a candidate
continuation action profile. Conditional on $\mathcal F_i$,
player $i$ chooses $a_i\in\mathbb R$ to maximise
\[
-E_i\left[(a_i-\theta({\bf k})-b_i)^2\right]
-\sigma_i\sum_{j\in N_i}E_i\left[(a_i-x_j)^2\right].
\]
Its second derivative with respect to $a_i$ is
$-2(1+\sigma_i n_i)<0$. Hence, the unique best response
is characterised by
\[
(1+\sigma_i n_i)a_i
=
\theta_i({\bf k})+b_i
+\sigma_i\sum_{j\in N_i}E_i(x_j).
\]
Consequently, $x$ is an equilibrium action profile if and only if,
for every $i\in N$,
\begin{equation}\label{eqn:operatorFOCi}
x_i
-
\frac{\sigma_i}{1+\sigma_i n_i}
\sum_{j\in N_i}E_i(x_j)
=
\frac{\theta_i({\bf k})+b_i}{1+\sigma_i n_i}
\qquad\text{almost surely}.
\end{equation}
Let $\boldsymbol\theta:=(\theta_i({\bf k}))_{i\in N}$ and
$b:=(b_i)_{i\in N}$. Stacking
\eqref{eqn:operatorFOCi} gives $(I-\mathcal G^N)x=L(\boldsymbol\theta+b)$, which can be 
written as 
\[
(I-\mathcal G^N)
\begin{bmatrix}
\tau_1^*(k_1,\widetilde{\beta}_1^*({\bf k}))\\
\vdots\\
\tau_n^*(k_n,\widetilde{\beta}_n^*({\bf k}))
\end{bmatrix}
=
D
\begin{bmatrix}
\theta_1(\mathbf{k})+b_1\\
\vdots\\
\theta_n(\mathbf{k})+b_n
\end{bmatrix}.
\]
Since conditional expectation is a contraction in $L^2$,
for every $x\in\mathcal H$,
\[
\begin{aligned}
\|(\mathcal G^N x)_i\|_2
&\leq
\frac{\sigma_i}{1+\sigma_i n_i}
\sum_{j\in N_i}\|E_i(x_j)\|_2\\
&\leq
\frac{\sigma_i n_i}{1+\sigma_i n_i}\|x\|.
\end{aligned}
\]
Therefore,
\[
\|\mathcal G^N\|
\leq
\max_{i\in N}
\frac{\sigma_i n_i}{1+\sigma_i n_i}
<1.
\]
It follows that $I-\mathcal G^N$ is invertible on
$\mathcal H$, with
\[
(I-\mathcal G^N)^{-1}
=
\sum_{r=0}^{\infty}(\mathcal G^N)^r,
\]
where the series converges in operator norm.
Since $\theta({\bf k})\in[0,n]$, the vector
$\boldsymbol\theta+b$ belongs to $\mathcal H$. Thus,
the unique solution is
\begin{eqnarray}\label{D}
x^*
=
(I-\mathcal G^N)^{-1}L(\boldsymbol\theta+b)
=
\mathcal C(\boldsymbol\theta+b).
\end{eqnarray}
Strict concavity ensures that this solution constitutes an
equilibrium action profile. From (\ref{D}), it follows that
\[
\begin{bmatrix}
\tau_1^*(k_1,\widetilde{\beta}_1^*({\bf k}))\\
\vdots\\
\tau_n^*(k_n,\widetilde{\beta}_n^*({\bf k}))
\end{bmatrix}
=
\begin{pmatrix}
\mathcal C_{11} & \mathcal C_{12} & \cdots & \mathcal C_{1n}\\
\mathcal C_{21} & \mathcal C_{22} & \cdots & \mathcal C_{2n}\\
\vdots & \vdots & \ddots & \vdots\\
\mathcal C_{n1} & \mathcal C_{n2} & \cdots & \mathcal C_{nn}
\end{pmatrix}.
\begin{bmatrix}
\theta_1(\mathbf{k})+b_1\\
\vdots\\
\theta_n(\mathbf{k})+b_n
\end{bmatrix}.
\]
Taking its $i$-th component yields
\eqref{eqn:equilibriumactionoperator}.
\end{proof}

The uniqueness result holds almost surely under the distribution of
histories induced by $\beta^*$; actions at histories outside equilibrium
play depend on the specified beliefs.

\begin{remark}
Theorem~\ref{thm:equilibriumaction} shows how network interactions
and private information jointly determine equilibrium actions.
Without network interactions, player $i$ chooses her expected bliss
point, $\theta_i({\bf k})+b_i$. With network interactions, her action
incorporates the expected bliss points of players in her 
component through the operators $\mathcal C_{ij}$. The expansion
$\mathcal C=\sum_{r=0}^{\infty}(\mathcal G^N)^rL$ captures direct
and indirect interactions, with successive conditional expectations
reflecting players' beliefs about others' information and actions.
Under complete information, these operators reduce to numerical
weights; when $\sigma_i=1$ for every $i$, the resulting matrix is
$C(1,G^N)$.
\end{remark}

\section{Characterisation of Equilibrium Message Strategy}

We distinguish between two types of neighbours, Bayesian neighbour and neive neoghbour, beginning with
Bayesian neighbours.

\subsection{Partition Equilibria under Public Communication}
\label{sec:public-partitions}

Our analysis draws on the argument of
Crawford and Sobel (1982). From Theorem 1 of Crawford and Sobel (1982) we get the existence of partition equilibrium. Consider communication from player $i$
to player $j$. Consider a complete network with player set
$N=\{1,\cdots,n\}$, where $n\geq2$.
The private types $k_i$, $i\in N$, are independently
and uniformly distributed on $[0,1]$, and
\[
\theta=\sum_{\ell\in N}k_\ell.
\]
After observing her type, each player sends a public
message. Messages are sent simultaneously and observed
by every player before actions are chosen.
Player $i$ then chooses $a_i\in\mathbb R$ and receives
utility
\[
u_i
=
-(a_i-\theta-b_i)^2
-\sum_{j\neq i}(a_i-a_j)^2.
\]
Thus, the coordination parameter is $\sigma_i=1$
for every player.

We consider pure reporting strategies and
square-integrable continuation actions.
Each player's message space is assumed to contain
enough distinct messages for the finite partitions
considered below.

For the complete network, the numerical centrality
matrix is
\[
C=\frac{I+J}{n+1},
\]
where $J$ is the $n\times n$ matrix of ones. In view of Theorem 4.4 we define the bias component of player $i$'s equilibrium action by
\[
B_i
:=
\sum_{\ell\in N}c_{i\ell}b_\ell
=
\frac{b_i+\sum_{\ell\in N}b_\ell}{n+1}.
\]
These constants satisfy
\begin{equation}\label{eq:public-bias-identities}
\sum_{\ell\in N}B_\ell=\sum_{\ell\in N}b_\ell,
\qquad
nB_i=b_i+\sum_{j\neq i}B_j,
\qquad
B_i-B_j=\frac{b_i-b_j}{n+1}.
\end{equation}

\begin{theorem}\label{thm:complete-network-information}
Consider a complete network with $n\geq2$ players,
independent types $k_i\sim\mathrm{Uniform}[0,1]$,
simultaneous public communication, and $\sigma_i=1$
for every player. Suppose that message spaces contain
enough distinct messages to represent any finite partition.

Let $N_i^{\max}(b_1,\ldots,b_n)$ denote the maximum
number of positive-length intervals in player $i$'s
equilibrium communication partition, with
$N_i^{\max}=\infty$ meaning that arbitrarily many
finite intervals are admissible. Then
\[
N_i^{\max}(b_1,\ldots,b_n)
=
\begin{cases}
\displaystyle
\left\lceil
\frac12+
\frac12\sqrt{
1+\frac{2(n-1)(n+1)}
{n\left|\displaystyle\sum_{j\neq i}(b_i-b_j)\right|}
}
\right\rceil-1,
&
\displaystyle\sum_{j\neq i}(b_i-b_j)\neq0,
\\[1.5em]
\infty,
&
\displaystyle\sum_{j\neq i}(b_i-b_j)=0.
\end{cases}
\]
Consequently, for every $b\in\mathbb R$,
\[
N_i^{\max}(b_1,\ldots,b_n)\longrightarrow\infty
\qquad\text{for every }i\in N
\]
as $(b_1,\ldots,b_n)\to(b,\ldots,b)$.
\end{theorem}

\begin{proof}
We establish the result for player $1$; the argument
for every other player follows by relabeling.

For each $i\in N$, define the bias component of her
continuation action by
\[
B_i
:=
\sum_{\ell=1}^{n}c'_{i\ell}b_\ell
=
\frac{b_i+\sum_{\ell=1}^{n}b_\ell}{n+1},
\]
where $c'_{ii}=2/(n+1)$ and $c'_{ij}=1/(n+1)$ for
$j\neq i$. These constants satisfy
\begin{equation}\label{eq:bias-identities}
b_i+\sum_{\ell\neq i}B_\ell=nB_i,
\qquad
B_i-B_j=\frac{b_i-b_j}{n+1}.
\end{equation}

\medskip
\noindent
\textit{Continuation actions.}
Fix a profile of pure public communication strategies.
For each player $i$, let
\[
q_i:=\mathbb E[k_i\mid m_i]
\]
denote the posterior mean induced by her public message.
Since each message depends only on its sender's type,
types remain independent conditional on the public
message profile ${\bf m}=(m_1,\ldots,m_n)$.

Write $\mathbb E_i$ for expectation conditional on
$(k_i,{\bf m})$. Player $i$'s action-stage first-order
condition is
\begin{equation}\label{eq:complete-best-response}
na_i
=
\mathbb E_i[\theta]+b_i
+\sum_{j\neq i}\mathbb E_i[a_j].
\end{equation}
Define
\[
\widehat a_i:=\mathbb E[a_i\mid{\bf m}].
\]
Taking expectations in
\eqref{eq:complete-best-response} conditional on
${\bf m}$ gives
\[
n\widehat a_i
=
\sum_{\ell=1}^{n}q_\ell+b_i
+\sum_{j\neq i}\widehat a_j.
\]
Equivalently,
\[
(n+1)\widehat a_i
=
\sum_{\ell=1}^{n}q_\ell+b_i
+\sum_{j=1}^{n}\widehat a_j.
\]
Summing over $i$ yields
\[
\sum_{j=1}^{n}\widehat a_j
=
n\sum_{\ell=1}^{n}q_\ell+\sum_{\ell=1}^{n}b_\ell.
\]
Substitution therefore gives
\begin{equation}\label{eq:expected-actions}
\widehat a_i
=
\sum_{\ell=1}^{n}q_\ell+B_i.
\end{equation}

Conditional on ${\bf m}$, player $i$'s private type
provides no additional information about another
player's type or prescribed action. Hence,
\[
\mathbb E_i[a_j]=\widehat a_j
\quad\text{for }j\neq i,
\qquad
\mathbb E_i[\theta]
=
k_i+\sum_{\ell\neq i}q_\ell.
\]
Using \eqref{eq:bias-identities} and
\eqref{eq:expected-actions} in
\eqref{eq:complete-best-response}, we obtain
\begin{equation}\label{eq:5.5}
\begin{aligned}
na_i
&=
k_i+\sum_{\ell\neq i}q_\ell+b_i
+\sum_{j\neq i}
\left(\sum_{\ell=1}^{n}q_\ell+B_j\right)\\
&=
k_i-q_i+n\sum_{\ell=1}^{n}q_\ell+nB_i.
\end{aligned}
\end{equation}
Thus,
\begin{equation}\label{eq:complete-actions}
a_i
=
\sum_{\ell=1}^{n}q_\ell
+\frac{k_i-q_i}{n}+B_i.
\end{equation}
These actions satisfy the best-response equations.
The conditional objective is strictly concave in
$a_i$, with second derivative $-2n$. Moreover,
the best-response mapping is a contraction with
modulus at most $(n-1)/n$ under the maximum component
$L^2$ norm. Therefore,
\eqref{eq:complete-actions} is the unique
square-integrable continuation equilibrium, up to
almost-sure equality.

\medskip
\noindent
\textit{Comparison of adjacent reports.}
Let $s:=k_1$, and consider a candidate partition
\[
0=p_0<p_1<\cdots<p_R=1.
\]
For an interior cutoff $p_r$, define the posterior
means of the adjacent intervals by
\[
q^+:=\frac{p_r+p_{r+1}}2,
\qquad
q^-:=\frac{p_{r-1}+p_r}2,
\qquad r=1,\ldots,R-1.
\]
Superscripts $+$ and $-$ indicate the upper- and
lower-interval reports, respectively.
Hold the other public reports fixed and write
\[
S:=\sum_{j=2}^{n}q_j.
\]

Holding the other players' continuation strategies
fixed, \eqref{eq:complete-actions} gives
\[
a_j^\pm
=
q^\pm+S+\frac{k_j-q_j}{n}+B_j,
\qquad j\neq1.
\]
Player $1$ retains knowledge of her actual type $s$,
so
\[
\mathbb E_1[\theta]=s+S,
\qquad
\mathbb E_1[a_j^\pm]=q^\pm+S+B_j.
\]
Her optimal continuation action satisfies
\[
\begin{aligned}
na_1^\pm
&=
s+S+b_1+\sum_{j=2}^{n}(q^\pm+S+B_j)\\
&=
s+(n-1)q^\pm+nS+nB_1.
\end{aligned}
\]
Therefore,
\begin{equation}\label{eq:5.7}
a_1^\pm
=
\frac{s}{n}+\frac{n-1}{n}q^\pm+S+B_1.
\end{equation}

For the payoff comparison, define
\[
T:=s+S+b_1,
\qquad
v_j:=\operatorname{Var}(k_j\mid m_j),
\]
and
\begin{equation}\label{eq:5.8}
A_j^\pm:=\mathbb E_1[a_j^\pm]=q^\pm+S+B_j.
\end{equation}
Here, $T$ is player $1$'s expected ideal action,
and $A_j^\pm$ is her expectation of player $j$'s
action under the corresponding report.

Conditional independence implies
\[
\operatorname{Var}_1(\theta)=\sum_{j=2}^{n}v_j,
\qquad
\operatorname{Var}_1(a_j^\pm)=\frac{v_j}{n^2}.
\]
Consequently,
\[
\mathbb E_1[u_1^\pm]
=
-(a_1^\pm-T)^2
-\sum_{j=2}^{n}(a_1^\pm-A_j^\pm)^2
-\left(1+\frac1{n^2}\right)\sum_{j=2}^{n}v_j.
\]
The variance terms are independent of player $1$'s
report. Indifference between the adjacent reports
therefore requires
\begin{equation}\label{eq:indifference}
\begin{aligned}
0={}&
(a_1^+-T)^2-(a_1^--T)^2\\
&+\sum_{j=2}^{n}
\left[
(a_1^+-A_j^+)^2-(a_1^--A_j^-)^2
\right].
\end{aligned}
\end{equation}

Write
\[
\Delta q:=q^+-q^->0,
\qquad
R_q:=q^++q^-.
\]
Substituting \eqref{eq:5.7} and \eqref{eq:5.8}
into \eqref{eq:indifference}, applying the
difference-of-squares identity, and dividing by
$\Delta q$ gives
\begin{equation}\label{eq:intermediate}
0
=
\frac{n-1}{n}(R_q-2s)
+\frac{2(n-1)}n(B_1-b_1)
-\frac2n\sum_{j=2}^{n}(B_1-B_j).
\end{equation}
Since
\[
\sum_{j=2}^{n}(B_1-B_j)
=
\frac1{n+1}\sum_{j=2}^{n}(b_1-b_j)
=
b_1-B_1,
\]
equation~\eqref{eq:intermediate} simplifies to
\[
0
=
\frac{n-1}{n}(R_q-2s)-2(b_1-B_1).
\]
At $s=p_r$,
\[
R_q-2s
=
\frac{p_{r+1}-2p_r+p_{r-1}}2.
\]
Hence the partition cutoffs satisfy
\begin{equation}\label{eq:final-difference}
p_{r+1}-2p_r+p_{r-1}
=
\frac{4n}{(n-1)(n+1)}
\sum_{j=2}^{n}(b_1-b_j),
\qquad r=1,\ldots,R-1.
\end{equation}
The comparison is independent of the other players'
realized reports. It therefore also applies at the
communication stage, before those reports are observed.

\medskip
\noindent
\textit{Admissible partitions.}
Successive interval lengths satisfy
\[
p_r-p_{r-1}
=
p_1+
\frac{4n}{(n-1)(n+1)}
\left[\sum_{j=2}^{n}(b_1-b_j)\right](r-1).
\]
Summing gives
\[
p_r
=
rp_1+
\frac{2n}{(n-1)(n+1)}
\left[\sum_{j=2}^{n}(b_1-b_j)\right]r(r-1).
\]
Imposing $p_R=1$ yields
\begin{equation}\label{eq:complete-first-cutoff}
p_1
=
\frac1R-
\frac{2n}{(n-1)(n+1)}
\left[\sum_{j=2}^{n}(b_1-b_j)\right](R-1).
\end{equation}
Thus,
\[
p_r
=
\frac rR+
\frac{2n}{(n-1)(n+1)}
\left[\sum_{j=2}^{n}(b_1-b_j)\right]r(r-R),
\]
and
\begin{equation}\label{eq:complete-interval-lengths}
p_r-p_{r-1}
=
\frac1R+
\frac{2n}{(n-1)(n+1)}
\left[\sum_{j=2}^{n}(b_1-b_j)\right](2r-R-1).
\end{equation}

If $\sum_{j=2}^{n}(b_1-b_j)>0$, interval lengths
increase with $r$. Consequently, all intervals have
positive length if and only if the first does:
\[
p_1
=
\frac1R-
\frac{2n}{(n-1)(n+1)}
\left[\sum_{j=2}^{n}(b_1-b_j)\right](R-1)>0.
\]

If $\sum_{j=2}^{n}(b_1-b_j)<0$, interval lengths
decrease with $r$. Positivity of the last interval
is therefore necessary and sufficient:
\[
1-p_{R-1}
=
\frac1R+
\frac{2n}{(n-1)(n+1)}
\left[\sum_{j=2}^{n}(b_1-b_j)\right](R-1)>0.
\]

In both nonzero cases, the shortest interval has
length
\[
\frac1R-
\frac{2n}{(n-1)(n+1)}
\left|\sum_{j=2}^{n}(b_1-b_j)\right|(R-1).
\]
Accordingly, admissibility is equivalent to
\[
R(R-1)
<
\frac{(n-1)(n+1)}
{2n\left|\displaystyle\sum_{j=2}^{n}(b_1-b_j)\right|}.
\]
Solving this inequality gives
\[
R<
\frac12+
\frac12\sqrt{
1+\frac{2(n-1)(n+1)}
{n\left|\displaystyle\sum_{j=2}^{n}(b_1-b_j)\right|}
}.
\]
Therefore, the largest admissible integer is
\[
N_1^{\max}
=
\left\lceil
\frac12+
\frac12\sqrt{
1+\frac{2(n-1)(n+1)}
{n\left|\displaystyle\sum_{j=2}^{n}(b_1-b_j)\right|}
}
\right\rceil-1.
\]
The ceiling minus one accounts for strict positivity
of interval lengths. The absolute value surrounds
the signed sum because its sign determines whether
the first or last interval is shortest.

If $\sum_{j=2}^{n}(b_1-b_j)=0$, the recurrence instead
gives
\[
p_r=\frac rR.
\]
Every finite $R$ is admissible, so $N_1^{\max}=\infty$. Finally, as $(b_1,\ldots,b_n)\to(b,\ldots,b)$,
\[
\left|\sum_{j=2}^{n}(b_1-b_j)\right|\longrightarrow0.
\]
For every fixed positive integer $R$, the
interval-positivity condition therefore holds
for all sufficiently close bias profiles.
This also holds when the bias sum is exactly zero.
It follows that $N_1^{\max}\to\infty$.
Relabeling establishes the result for every player.
\end{proof}

\begin{remark} The point to note here is that the number of sub-interval of the type space in our case is weakly greater then that of Crawford Sobel 1982. Thus, partitioning the type space, our case is a refinement, implying more informative cheap-talk. Moreover, for non-negativity, we can assume $b_1>b_2.$ For player 2 the reverse will happen. Since everything is symmetric in the game (except biases), one can intuitively think $N_{1}(b_1, b_2)=N_{2}(b_1, b_2)$ because the absolute distance between the biases is what matters. 
\end{remark}

We can similarly prove, when $N=\{1,2,3\}$ with $g_{12}=g_{13}=g_{23}=1$ and public communication, $N_{1}(b_1, b_2, b_3) = [\frac{1}{2} + \frac{1}{2}(1 + \frac{16/3}{(b_1 - b_2) + (b_1 - b_3)})^{\frac{1}{2}}].$ Due to public communication if $b_1$ and $b_3$ are far apart, almost full informative communication will not happen between $1$ and $2$ even if $b_1 \rightarrow b_2.$

In case of two players, maximum number of subinterval is 
$
[\frac{1}{2} + \frac{1}{2}\sqrt{(1+\frac{3}{b_1 - b_2})}], 
$
which is greater than 
$
[\frac{1}{2} + \frac{1}{2}\sqrt{(1+\frac{2}{b})}], 
$
the maximum number of subinterval for CS. Now if we have three players and complete network, then the maximum number of sub-interval for player 1, $N_{1}(b_1, b_2, b_3)$ is,
\[
\left[\frac{1}{2} + \frac{1}{2}(1 + \frac{6}{(b_1 - b_2) + (b_1 - b_3)})^{\frac{1}{2}}\right].
\]
If we have a start network with player 2 being the centre, then $N_{1}(b_1, b_2, b_3)= [\frac{\log x^{*}}{\log \lambda}],
$
where, 
\[
x^{*}=\frac{1}{2}[(\lambda+1)\left(1+\frac{1}{M}\right)+\sqrt{(\lambda+1)^2\left(1+\frac{1}{M}\right)^2 -4\lambda}
\]
where,
\[
M_1:=\frac{5(3b_1-2b_2-b_3)-4}{8}>0,
\qquad
\lambda:=\frac{7+2\sqrt{10}}{3}.
\] 
Calculation of 3 player star network is given in the \textit{Appendix}.

\subsection{Communication with a Naive Neighbour}

We now consider communication from player $i$ to a naive
neighbour $k\in N_i$. Player $k$ takes the report
$m_{ik}$ at face value, interpreting it as the true value
of player $i$'s private information. Player $i$, however,
knows her true type and anticipates how her report affects
the recipient's action.

We introduce a cost of misreporting and assume a common
coordination parameter $\sigma>0$. Suppose $\mathcal N_i \subseteq N_i$ is the set of naive neighbours of player $i$ and $\mathcal N_i \neq \phi$. Player $i$'s new utility is defined as  $\widetilde{u}_{i}: \Theta \times \prod\limits_{j \in N_{i} \cup \{i\}}A_{j} \times K_i \times \prod\limits_{j \in \mathcal N_i}M_{ij}  \to \mathbb{R}$, and assumed to have the following form.
\begin{equation}\label{eqn:utility-naive}
\begin{aligned}
u_i\bigl(\theta,(a_j)_{j\in N_i\cup\{i\}},k_i,(m_{ij})_{j \in \mathcal N_i}\bigr)
={}&
-\bigl(a_i-\theta({\bf k})-b_i\bigr)^2\\
&-\sigma\sum_{j\in N_i}(a_i-a_j)^2
-\gamma\sum_{j\in \mathcal N_i}|m_{ij}-k_i|,
\end{aligned}
\end{equation}
where $\gamma\geq0$ measures the marginal cost of
misreporting. The cost is symmetric: over-reporting
and under-reporting by the same amount incur the
same penalty. We first study player $i$'s conditional reporting
incentive, holding her other outgoing messages and
her incoming messages fixed.

The recipient's response depends on her beliefs
about both the state and other players' actions.
We therefore state the result for a given affine
continuation response, whose coefficient will be
characterised separately. All continuation actions
are assumed to be square-integrable.

\begin{proposition}
\label{prop:naive-report-direction}
Fix player $i$, a naive neighbour $k\in N_i$, and
a true type $\overline{k}_i\in(0,1)$.  Write $m:=m_{ik}$.
Hold all other messages, including the message
received from $k$, and the other players'
continuation strategies fixed. Let $\overline{\tau}_k$ denote the recipient's
continuation action under the truthful report
$m=\overline{k}_i$. Suppose that, for every
$m\in[0,1]$,
\[
\tau_k(m)
=
\overline{\tau}_k+w_{ik}(m-\overline{k}_i),
\qquad w_{ik}>0,
\]
where $w_{ik}$ is deterministic and independent of
$m$. Suppose also that the prescribed actions of
players in $N_i\setminus\{k\}$ are unaffected by $m$.
Player $i$ chooses her continuation action optimally
after reporting. Let $\overline{\tau}_i$ denote player $i$'s optimal
continuation action under the truthful report, and
define
\[
\overline y_k:=\mathbb E_i[\overline{\tau}_k].
\]
Then player $i$'s conditional expected utility has
a unique maximizer $m_{ik}^*\in[0,1]$, satisfying
\[
\begin{aligned}
m_{ik}^*>\overline{k}_i
&\quad\Longleftrightarrow\quad
\overline{\tau}_i-\overline y_k
>\frac{\gamma}{2\sigma w_{ik}},
\\[4pt]
m_{ik}^*=\overline{k}_i
&\quad\Longleftrightarrow\quad
\left|\overline{\tau}_i-\overline y_k\right|
\leq\frac{\gamma}{2\sigma w_{ik}},
\\[4pt]
m_{ik}^*<\overline{k}_i
&\quad\Longleftrightarrow\quad
\overline{\tau}_i-\overline y_k
<-\frac{\gamma}{2\sigma w_{ik}}.
\end{aligned}
\]
\end{proposition}

In first case $(m_{ik}^*>\overline{k}_i)$ it is assumed that $\overline{k}_i<1.$ Otherwise analysis of over-reporting does not make any sense. Same for the case $(m_{ik}^*<\overline{k}_i)$ we must have $\overline{k}_i>0.$ The proposition compares the sender's optimal action
under truthful reporting with her expectation of the
recipient's action. An upward report raises the
recipient's action and is attractive when that action
is sufficiently below the sender's. A downward report
is attractive when the reverse holds. Truthful
reporting is optimal when the expected coordination
gain from a marginal distortion does not exceed its
cost. Holding the truthful action difference fixed,
a larger $\gamma$ widens this truth-telling region,
whereas a larger response coefficient $w_{ik}$
narrows it. Larger $w_{ik}$ implies player $k$ cares more about $i$ and vice-versa. Thus even if misreporting is costly, $i$ will misreport to $k$. Similarly smaller $w_{ik}$ makes the tradeoff of misreporting high.

\begin{proof}

Define $\theta_i:=\mathbb E_i[\theta],$ $y_j:=\mathbb E_i[\tau_j], \ j\in N_i\setminus\{k\},$ and $y_k(m):=\mathbb E_i[\tau_k(m)].$
Thus, $y_j$ and $y_k(m)$ are player $i$'s expectations
of her neighbours' actions. By assumption, $\theta_i$
and $y_j$, $j\neq k$, are independent of $m$, while
\[
y_k(m)
=
\overline y_k+w_{ik}(m-\overline{k}_i),
\qquad
\overline y_k:=\mathbb E_i[\overline{\tau}_k].
\]

Player $i$'s continuation payoff is strictly concave
in her own action. Her optimal action therefore
satisfies
\begin{equation}\label{eq:sender-action-foc}
(1+\sigma n_i)\tau_i(m)
=
\theta_i+b_i
+\sigma\sum_{j\in N_i\setminus\{k\}}y_j
+\sigma y_k(m).
\end{equation}
Consequently,
\begin{equation}\label{eq:action-derivatives}
\tau_i'(m)=\frac{\sigma w_{ik}}{1+\sigma n_i},
\qquad
y_k'(m)=w_{ik},
\end{equation}
and hence
\begin{equation}\label{eq:difference-derivative}
\frac{d}{dm}\bigl(\tau_i(m)-y_k(m)\bigr)
=
-\frac{1+\sigma(n_i-1)}{1+\sigma n_i}w_{ik}.
\end{equation}

Up to an additive term independent of $m$, player
$i$'s conditional expected payoff is
\[
U_i(m)=V_i(m)-\gamma|m-\overline{k}_i|,
\]
where
\begin{align}
V_i(m)
={}&
-\bigl(\tau_i(m)-\theta_i-b_i\bigr)^2
-\sigma\sum_{j\in N_i\setminus\{k\}}
\bigl(\tau_i(m)-y_j\bigr)^2
\nonumber\\
&-\sigma\bigl(\tau_i(m)-y_k(m)\bigr)^2.
\label{eq:smooth-report-payoff}
\end{align}
The omitted terms comprise conditional variances
and the costs of the other reports. They are
independent of $m$, since the recipient's action
changes by a deterministic additive shift. Differentiating \eqref{eq:smooth-report-payoff} gives
\begin{align*}
V_i'(m)
={}&
-2\tau_i'(m)
\left[
\tau_i(m)-\theta_i-b_i
+\sigma\sum_{j\in N_i\setminus\{k\}}
\bigl(\tau_i(m)-y_j\bigr)
+\sigma\bigl(\tau_i(m)-y_k(m)\bigr)
\right]\\
&+
2\sigma w_{ik}\bigl(\tau_i(m)-y_k(m)\bigr).
\end{align*}
The bracketed term vanishes by
\eqref{eq:sender-action-foc}. Thus,
\begin{equation}\label{eq:smooth-payoff-derivative}
V_i'(m)
=
2\sigma w_{ik}\bigl(\tau_i(m)-y_k(m)\bigr).
\end{equation}
Using \eqref{eq:difference-derivative},
\[
V_i''(m)
=
-\frac{
2\sigma w_{ik}^{\,2}[1+\sigma(n_i-1)]
}{
1+\sigma n_i
}
<0.
\]
Since $-\gamma|m-\overline{k}_i|$ is concave,
$U_i$ is strictly concave and continuous on $[0,1]$.
It therefore admits a unique maximiser.

At the truthful report,
$\tau_i(\overline{k}_i)=\overline{\tau}_i$ and
$y_k(\overline{k}_i)=\overline y_k$.
Equation~\eqref{eq:smooth-payoff-derivative} therefore
implies
\begin{equation}\label{eq:one-sided-report-derivatives}
\begin{aligned}
U'_{i,+}(\overline{k}_i)
&=
2\sigma w_{ik}
(\overline{\tau}_i-\overline y_k)-\gamma,\\
U'_{i,-}(\overline{k}_i)
&=
2\sigma w_{ik}
(\overline{\tau}_i-\overline y_k)+\gamma.
\end{aligned}
\end{equation}
Because $\overline{k}_i$ is interior, strict concavity
implies that the maximiser lies above
$\overline{k}_i$ if and only if
$U'_{i,+}(\overline{k}_i)>0$, and below
$\overline{k}_i$ if and only if
$U'_{i,-}(\overline{k}_i)<0$.
It equals $\overline{k}_i$ if and only if
\[
U'_{i,-}(\overline{k}_i)\geq0
\quad\text{and}\quad
U'_{i,+}(\overline{k}_i)\leq0.
\]
Substituting \eqref{eq:one-sided-report-derivatives}
establishes the three stated conditions.
\end{proof}

\begin{proposition}[Optimal report and its response to private information]
\label{prop:optimal-naive-report}
Maintain the assumptions of
Proposition~\ref{prop:naive-report-direction},
but allow $\overline{k}_i\in[0,1]$. Define
\[
d_{ik}(\overline{k}_i)
:=
\overline{\tau}_i-\overline y_k.
\]
This is the difference between player $i$'s own action
and her expectation of player $k$'s action under
the truthful report. The following statements hold.
\begin{enumerate}
\item[(a)]
The unique optimal report is
\begin{equation}\label{eq:optimal-report}
m_{ik}^*
=
\begin{cases}
\displaystyle
\min\left\{
1,\,
\overline{k}_i+
\frac{1+\sigma n_i}
     {w_{ik}[1+\sigma(n_i-1)]}
\left[
d_{ik}(\overline{k}_i)
-\frac{\gamma}{2\sigma w_{ik}}
\right]
\right\},
&
\displaystyle
d_{ik}(\overline{k}_i)>
\frac{\gamma}{2\sigma w_{ik}},
\\[1.5em]
\overline{k}_i,
&
\displaystyle
\left|d_{ik}(\overline{k}_i)\right|
\leq\frac{\gamma}{2\sigma w_{ik}},
\\[1.5em]
\displaystyle
\max\left\{
0,\,
\overline{k}_i+
\frac{1+\sigma n_i}
     {w_{ik}[1+\sigma(n_i-1)]}
\left[
d_{ik}(\overline{k}_i)
+\frac{\gamma}{2\sigma w_{ik}}
\right]
\right\},
&
\displaystyle
d_{ik}(\overline{k}_i)<
-\frac{\gamma}{2\sigma w_{ik}}.
\end{cases}
\end{equation}

\item[(b)]
Suppose additionally that
$\theta=\sum_{\ell\in N}k_\ell$, private information is
independent across players, and the fixed incoming
messages convey no information about $k_i$.
Holding the report and all other messages fixed,
suppose that the conditional distribution of the
other players' actions does not depend on
$\overline{k}_i$. Assume also that $w_{ik}$ is
independent of $\overline{k}_i$.

Within either interior misreporting region,
\begin{equation}\label{eq:optimal-report-type-derivative}
\frac{\partial m_{ik}^*}{\partial\overline{k}_i}
=
\frac{1}{w_{ik}[1+\sigma(n_i-1)]}
>0.
\end{equation}
Within a truthful-reporting region, the derivative
equals $1$. Within a region where the optimal report
is constrained to $0$ or $1$, it equals $0$.
The optimal reporting rule is continuous and weakly
increasing in actual private information, and strictly
increasing on intervals where the optimal report
lies in $(0,1)$.
\end{enumerate}
\end{proposition}

\begin{proof}
We retain the notation from
Proposition~\ref{prop:naive-report-direction}.
Since the recipient's action is affine in the report,
\[
y_k(m)
=
\overline y_k+w_{ik}(m-\overline{k}_i).
\]
Integrating the sender's action response in
\eqref{eq:action-derivatives} gives
\begin{equation}\label{eq:sender-action-response}
\tau_i(m)
=
\overline{\tau}_i+
\frac{\sigma w_{ik}}{1+\sigma n_i}
(m-\overline{k}_i).
\end{equation}
Consequently,
\begin{equation}\label{eq:action-gap-response}
\tau_i(m)-y_k(m)
=
d_{ik}(\overline{k}_i)
-
\frac{w_{ik}[1+\sigma(n_i-1)]}{1+\sigma n_i}
(m-\overline{k}_i).
\end{equation}

\medskip
\noindent
\textit{Part (a).}
First consider an interior optimum satisfying
$\overline{k}_i<m_{ik}^*<1$.
In this region,
\[
U_i(m)=V_i(m)-\gamma(m-\overline{k}_i).
\]
By \eqref{eq:smooth-payoff-derivative},
the first-order condition is
\[
2\sigma w_{ik}
\bigl(\tau_i(m_{ik}^*)-y_k(m_{ik}^*)\bigr)
-\gamma=0.
\]
Substituting \eqref{eq:action-gap-response},
\[
2\sigma w_{ik}
\left[
d_{ik}(\overline{k}_i)
-
\frac{w_{ik}[1+\sigma(n_i-1)]}{1+\sigma n_i}
(m_{ik}^*-\overline{k}_i)
\right]
-\gamma=0.
\]
Rearranging gives
\[
\frac{w_{ik}[1+\sigma(n_i-1)]}{1+\sigma n_i}
(m_{ik}^*-\overline{k}_i)
=
d_{ik}(\overline{k}_i)
-\frac{\gamma}{2\sigma w_{ik}}.
\]
Therefore,
\begin{equation}\label{eq:optimal-report-over}
m_{ik}^*
=
\overline{k}_i+
\frac{1+\sigma n_i}
     {w_{ik}[1+\sigma(n_i-1)]}
\left[
d_{ik}(\overline{k}_i)
-\frac{\gamma}{2\sigma w_{ik}}
\right].
\end{equation}
When the expression on the right exceeds $1$,
strict concavity implies that the constrained
optimum is $1$.

For an interior optimum satisfying
$0<m_{ik}^*<\overline{k}_i$,
\[
U_i(m)=V_i(m)-\gamma(\overline{k}_i-m).
\]
The first-order condition becomes
\[
2\sigma w_{ik}
\bigl(\tau_i(m_{ik}^*)-y_k(m_{ik}^*)\bigr)
+\gamma=0.
\]
Substituting \eqref{eq:action-gap-response} and
rearranging yields
\begin{equation}\label{eq:optimal-report-under}
m_{ik}^*
=
\overline{k}_i+
\frac{1+\sigma n_i}
     {w_{ik}[1+\sigma(n_i-1)]}
\left[
d_{ik}(\overline{k}_i)
+\frac{\gamma}{2\sigma w_{ik}}
\right].
\end{equation}
If this expression is negative, the constrained
optimum is $0$.

When
\[
|d_{ik}(\overline{k}_i)|
\leq\frac{\gamma}{2\sigma w_{ik}},
\]
the one-sided derivatives in
\eqref{eq:one-sided-report-derivatives}
imply that truthful reporting is optimal.
Combining these cases with the bounds $0\leq m\leq1$
gives \eqref{eq:optimal-report}. The clipping at $0$
and $1$ also covers the endpoint types
$\overline{k}_i=0$ and $\overline{k}_i=1$.

\medskip
\noindent
\textit{Part (b).}
Vary $\overline{k}_i$ while holding all other messages
fixed. Under the additional assumptions,
\[
\frac{\partial\theta_i}{\partial\overline{k}_i}=1.
\]
At a fixed report, the conditional expectations of
the other players' actions do not depend on
$\overline{k}_i$. Along the truthful benchmark,
however, the report itself equals $\overline{k}_i$.
Hence,
\begin{equation}\label{eq:recipient-truth-type-response}
\frac{\partial\overline y_k}
     {\partial\overline{k}_i}
=w_{ik}.
\end{equation}

Evaluating \eqref{eq:sender-action-foc} at truth
and differentiating gives
\[
(1+\sigma n_i)
\frac{\partial\overline{\tau}_i}
     {\partial\overline{k}_i}
=
1+\sigma w_{ik}.
\]
Therefore,
\begin{equation}\label{eq:sender-truth-type-response}
\frac{\partial\overline{\tau}_i}
     {\partial\overline{k}_i}
=
\frac{1+\sigma w_{ik}}{1+\sigma n_i}.
\end{equation}
Subtracting
\eqref{eq:recipient-truth-type-response}
from \eqref{eq:sender-truth-type-response},
\begin{align}
d_{ik}'(\overline{k}_i)
&=
\frac{1+\sigma w_{ik}}{1+\sigma n_i}-w_{ik}
\nonumber\\
&=
\frac{1-w_{ik}[1+\sigma(n_i-1)]}
     {1+\sigma n_i}.
\label{eq:truthful-gap-type-response}
\end{align}

Differentiating either interior reporting formula,
\eqref{eq:optimal-report-over} or
\eqref{eq:optimal-report-under}, yields
\begin{align*}
\frac{\partial m_{ik}^*}{\partial\overline{k}_i}
&=
1+
\frac{1+\sigma n_i}
     {w_{ik}[1+\sigma(n_i-1)]}
d_{ik}'(\overline{k}_i)
\\
&=
1+
\frac{1-w_{ik}[1+\sigma(n_i-1)]}
     {w_{ik}[1+\sigma(n_i-1)]}
\\
&=
\frac{1}{w_{ik}[1+\sigma(n_i-1)]}
>0.
\end{align*}
Within a truthful-reporting region,
$m_{ik}^*=\overline{k}_i$, so the derivative equals $1$.
Within a region where a boundary constraint binds,
the report is constant and its derivative equals $0$.

At the thresholds between truthful reporting and
misreporting, the adjustment term in
\eqref{eq:optimal-report} vanishes. At a boundary
constraint, the interior expression meets $0$ or $1$.
The reporting rule is therefore continuous across
regions. Its slopes are nonnegative throughout
and strictly positive on its interior branches,
which establishes the stated monotonicity.
\end{proof}


\medskip
\noindent

\medskip
\noindent

\section{Endogenous Bias}\label{Sec:5}

Preferences need not be exogenously fixed. Becker and Mulligan (1997)
model the endogenous determination of time preference, allowing
individuals to devote resources to reducing their discounting of
future utility. Motivated by this perspective, we allow players to
choose their biases. In a network, a change in one player's bias
affects both her own action and the actions of other players,
potentially altering their incentives to choose their biases.
Bias choices are therefore strategically interdependent.
The following result characterizes how a change in one player's
bias affects the biases chosen by other players.
\vspace{0.2cm}

\begin{proposition}\label{prop:bias-response}
Let $\langle N,g,(b_i)_{i\in N}\rangle$ be a two-stage
communication and coordination game, and let
$C=[c_{ij}]_{i,j\in N}$ be the centrality matrix defined
above. Suppose that $k\in N_i$. Then the optimal bias
of player $i$ is increasing in the bias of player $k$;
that is,
\[
0<\frac{\partial b_i}{\partial b_k}<1.
\]
\end{proposition}

\begin{proof}
Differentiating the expected utility of player $i$
with respect to $b_i$ and from the first-order
condition, we get
\begin{equation}\label{eqn:1}
-(\tau_i-\theta_i(\mathbf{k})-b_i)(c_{ii}-1)
=
\sum_{j\in N_i}(\tau_i-\tau_j)(c_{ii}-c_{ji}).
\end{equation}
This gives the optimal choice of $b_i$. From this
we get
\[
-(c_{ii}-1)
\bigl[(c_{ii}-1)\,\partial b_i+c_{ik}\,\partial b_k\bigr]
=
\sum_{j\in N_i}
\bigl[(c_{ii}-c_{ji})\,\partial b_i
+(c_{ik}-c_{ji})\,\partial b_k\bigr]
(c_{ii}-c_{ji}).
\]
This implies
\[
\frac{\partial b_i}{\partial b_k}
=
\frac{
c_{ik}(1-c_{ii})
+\displaystyle\sum_{j\in N_i}
(c_{jk}-c_{ik})(c_{ii}-c_{ji})
}{
(c_{ii}-1)^2
+\displaystyle\sum_{j\in N_i}(c_{ii}-c_{ji})^2
}.
\]

We now simplify the numerator. Recall that
\[
(I-G^N)C=L.
\]
Taking the $(i,i)$-th entry gives
\[
c_{ii}
-\frac{1}{1+n_i}\sum_{j\in N_i}c_{ji}
=
\frac{1}{1+n_i}.
\]
Thus,
\[
(1+n_i)c_{ii}
=
1+\sum_{j\in N_i}c_{ji},
\]
and consequently
\[
1-c_{ii}
=
\sum_{j\in N_i}(c_{ii}-c_{ji}).
\]
Substituting this identity into the numerator yields
\[
\begin{aligned}
&c_{ik}(1-c_{ii})
+\sum_{j\in N_i}
(c_{jk}-c_{ik})(c_{ii}-c_{ji})\\
&\qquad=
c_{ik}\sum_{j\in N_i}(c_{ii}-c_{ji})
+\sum_{j\in N_i}
(c_{jk}-c_{ik})(c_{ii}-c_{ji})\\
&\qquad=
\sum_{j\in N_i}c_{jk}(c_{ii}-c_{ji}).
\end{aligned}
\]
Therefore,
\[
\frac{\partial b_i}{\partial b_k}
=
\frac{
\displaystyle\sum_{j\in N_i}
c_{jk}(c_{ii}-c_{ji})
}{
(c_{ii}-1)^2
+\displaystyle\sum_{j\in N_i}(c_{ii}-c_{ji})^2
}.
\]

By Lemma 3.2, we have $c_{ii}>c_{ji} \text{for every }j\in N_i,$ and all entries of $C$ are nonnegative. Hence,
\[
\sum_{j\in N_i}c_{jk}(c_{ii}-c_{ji})>0.
\]
The denominator is also strictly positive. It follows
that $\frac{\partial b_i}{\partial b_k}>0.$

Now let the numerator and denominator be denoted by
$A$ and $B$, respectively. Then,
\begin{align*}
B-A
&=
(1-c_{ii})^2
+\sum_{j\in N_i}(c_{ii}-c_{ji})^2
-\sum_{j\in N_i}c_{jk}(c_{ii}-c_{ji})\\
&=
\sum_{j\in N_i}
(c_{ii}-c_{ji})
\bigl[(1-c_{ii})+(c_{ii}-c_{ji})-c_{jk}\bigr]\\
&=
\sum_{j\in N_i}
(c_{ii}-c_{ji})(1-c_{ji}-c_{jk}).
\end{align*}
By Lemma 3.2, $c_{ii}>c_{ji}$ for all $j\neq i.$
Moreover,
\[
1-c_{ji}-c_{jk}
=
\sum_{\ell\in N_j\setminus\{i,k\}}c_{j\ell}
>0.
\]
Therefore,
\[
B-A
=
\sum_{j\in N_i}
(c_{ii}-c_{ji})
\sum_{\ell\in N\setminus\{i,k\}}c_{j\ell}
>0.
\]
Hence, $A<B$. Thus our result follows.
\end{proof}

 This shows the effect of change in bias of a player to change in another player's bias. Player $i$ observes how much player $k$ contributes to her ($i's$) neighbours' centrality compare to $k$'s influence on her own ($i$). If $k's$ influence is higher on every neighbour (of $i$),she changes her own bias in same direction with change in $k's$ bias. But if the exact opposite is the case, then $i$ may not go with same direction as $k$. This may happen for several reasons. Consider a political scenario and the players being political parties. When player $k$'s influence is the strongest on player $i$ (compare to neighbours of $i$), any change in $k$’s bias affects $i$’s strategic environment more than others. So if $k$ increases bias, $i$ may decrease its bias to avoid appearing too aligned or dependent, especially when $k$'s influence on others is weak, making $i$'s reaction highly visible. This opposite movementshelps $i$ preserve independence, relevance, and strategic balance. Also, $i$ observes $k$ is not much liked by others. So, going in same direction with $k$ may increase the distance of $i$ from others. 

But suppose for some neighbours of $i$, $k's$ influence is higher only to them than that of to player $i$. Then player $i$ adjusts the bias using how much player $i$ influences the respective players. Suppose $i$ can influence player $j$ much compare to another player $p$. Since this is known to $i,$she understands her own preference already affects $j$ up to some large extent compare to $p.$ But $p$ being less affected by $i,$ $p$ is acting more independently from $i's$ perspective. So, $i$ pays more attention to how $k$ influences $p$.

\section{Conclusion}\label{Sec:6}

The credibility of communication depends on the network through
which it takes place. Preference differences matter, but the extent of
their consequences are shaped by who can communicate with
whom, what others observe, and whose actions players wish
to influence. Network structure therefore affects not only
how information reaches decision makers, but also the
incentives to disclose or distort it. This connects strategic
information transmission with local coordination and
highlights why bilateral bias differences alone cannot
fully explain reporting behaviour.

A central implication is that aligning preferences and
improving access to information address different sources
of communication failure. Shared objectives can support
increasingly informative communication when reports are
publicly observed, while restricted access to others'
reports can create additional incentives for distortion.
Likewise, the incentive to mislead a naive recipient depends
on how strongly that recipient responds and how the sender
values the resulting change in actions. The same cost of
misreporting may therefore sustain truthfulness in one
relationship but fail to do so in another.

For organisations that rely on decentralised information,
these findings suggest that reporting incentives should
be considered alongside the allocation of responsibilities
and communication channels. Reducing preference conflict
alone may leave informational sources of disagreement
unresolved. Changes in transparency or connectivity must
also be assessed through their effects on beliefs and
coordination incentives. Understanding these effects when
communication relationships are chosen endogenously or
develop through repeated interaction is a natural direction
for further research.

\section{Appendix}

\textbf{Continuation Actions and Partition Cutoffs on the Three-Player Star}

Consider the network
\[
g_{12}=g_{21}=g_{23}=g_{32}=1,
\qquad
g_{13}=g_{31}=0.
\]
Player $2$ is the centre, and players $1$ and $3$
are the leaves. Types $k_1,k_2,k_3$ are independently
and uniformly distributed on $[0,1]$, and
\[
\theta=k_1+k_2+k_3.
\]
Player $i$ chooses $a_i\in\mathbb R$ and receives utility
\[
u_i=-(a_i-\theta-b_i)^2
-\sum_{j\in N_i}(a_i-a_j)^2.
\]

Communication is simultaneous and precedes the choice
of actions. Player $2$ sends the same message $m_2$
to both leaves. Player $1$ sends $m_1$ only to player
$2$, and player $3$ sends $m_3$ only to player $2$.
Neither leaf observes the other leaf's message,
and messages are not forwarded.

We consider pure communication strategies. Let
\[
q_i:=\mathbb E[k_i\mid m_i]
\]
denote the posterior mean induced by player $i$'s
message. Each player knows her own type, her outgoing
message, and her incoming messages. Thus, at the
action stage,
\[
\begin{aligned}
E_1[X]&:=\mathbb E[X\mid k_1,m_1,m_2],\\
E_2[X]&:=\mathbb E[X\mid k_2,m_1,m_2,m_3],\\
E_3[X]&:=\mathbb E[X\mid k_3,m_2,m_3].
\end{aligned}
\]
All conditional-expectation identities for a fixed strategy
profile are understood almost surely under its induced
distribution. When comparing reports, we hold the other
players' continuation strategies fixed and allow the sender
to choose her optimal continuation action.

For player $i$, write a partition with $R_i$ intervals as
\[
0=p_{i,0}<p_{i,1}<\cdots<p_{i,R_i}=1.
\]
Under the uniform prior, the posterior mean associated
with interval $r$ is
\[
\frac{p_{i,r-1}+p_{i,r}}2.
\]
We assume that message spaces contain enough distinct
messages to represent the partitions under consideration.
An admissible cutoff solution has strictly positive interval
lengths. For player $2$, the conditional cutoff system below
is evaluated at fixed incoming posterior means; it is not
a communication-stage equilibrium characterization for the
simultaneous-message game.

\begin{proposition}\label{prop:star-partitions}
Consider the three-player star network with player $2$ as
the centre. The following statements hold.

\begin{enumerate}
\item[\emph{(i)}]
For a fixed pure communication strategy profile under the
message-observation structure specified above, the unique
square-integrable continuation equilibrium actions satisfy,
almost surely,
\begin{equation}\label{eq:star-actions}
\begin{aligned}
a_1
&=
\frac12 k_1+\frac3{10}q_1+q_2+\frac35
+\frac{5b_1+2b_2+b_3}{8},
\\
a_2
&=
\frac13 k_2+\frac23q_2+\frac35(q_1+q_3)+\frac25
+\frac{b_1+2b_2+b_3}{4},
\\
a_3
&=
\frac12 k_3+\frac3{10}q_3+q_2+\frac35
+\frac{b_1+2b_2+5b_3}{8}.
\end{aligned}
\end{equation}

\item[\emph{(ii)}]
Suppose $3b_1-2b_2-b_3>4/5$. Player $1$'s partition
cutoffs satisfy
\begin{equation}\label{eq:leaf-recurrence}
3p_{1,r+1}-14p_{1,r}+3p_{1,r-1}
=5(3b_1-2b_2-b_3)-4,
\quad r=1,\ldots,R_1-1.
\end{equation}
Define
\[
M_1:=\frac{5(3b_1-2b_2-b_3)-4}{8}>0,
\qquad
\lambda:=\frac{7+2\sqrt{10}}{3},
\]
and
\[
x^*
:=
\frac{
\left(1+\frac1{M_1}\right)(\lambda+1)
+
\sqrt{
\left(1+\frac1{M_1}\right)^2(\lambda+1)^2
-4\lambda
}
}{2}.
\]
The maximum number of positive-length intervals satisfying
player $1$'s reporting incentives is
\[
R_1^{\max}
=
\left\lceil
\frac{\log x^*}{\log\lambda}
\right\rceil-1.
\]
The corresponding characterization for player $3$ follows
by interchanging players $1$ and $3$, provided that
$3b_3-2b_2-b_1>4/5$.

\item[\emph{(iii)}]
For player $2$'s conditional indifference calculation,
hold the incoming posterior means
\[
P:=\mathbb E[k_1\mid m_1],
\qquad
Q:=\mathbb E[k_3\mid m_3]
\]
fixed, and suppose
\[
\frac{12}{5}(P+Q-1)+\frac32(2b_2-b_1-b_3)>0.
\]
The conditional cutoff equation is
\[
p_{2,r+1}-2p_{2,r}+p_{2,r-1}
=\frac{12}{5}(P+Q-1)+\frac32(2b_2-b_1-b_3),
\quad r=1,\ldots,R_2-1.
\]
With $p_{2,0}=0$ and $p_{2,R_2}=1$, the maximum number
of positive-length intervals solving this conditional system is
\begin{equation}\label{eq:centre-maximum}
R_2^{\max}(P,Q)
=
\left\lceil
\frac{
1+\sqrt{
1+\dfrac{80}{
24(P+Q-1)+15(2b_2-b_1-b_3)
}
}
}{2}
\right\rceil-1.
\end{equation}
This conditional count holds with the continuation strategies
used in the comparison fixed. It does not assert that player
$2$ can condition her outgoing message on $P,Q$ in the
simultaneous-communication game.
\end{enumerate}
\end{proposition}
\begin{proof}
We derive the continuation actions directly from the
three best-response equations. We then compare
adjacent reports and calculate the first cutoff
and maximum number of intervals for each relevant
player.

\medskip
\noindent
\textit{1. Expected states and best responses.}

Independence of private information implies
\[
E_1[\theta]=k_1+q_2+\frac12,
\qquad
E_2[\theta]=q_1+k_2+q_3,
\qquad
E_3[\theta]=\frac12+q_2+k_3.
\]
The action-stage first-order conditions are
\begin{equation}\label{eq:action-focs}
\begin{aligned}
2a_1
&=k_1+q_2+\frac12+b_1+E_1[a_2],\\
3a_2
&=q_1+k_2+q_3+b_2+E_2[a_1]+E_2[a_3],\\
2a_3
&=\frac12+q_2+k_3+b_3+E_3[a_2].
\end{aligned}
\end{equation}
These equations must be solved jointly.

\medskip
\noindent
\textit{2. Action expectations conditional only on
the centre's message.}

Because $m_2$ depends only on $k_2$,
\[
\mathbb E[\theta\mid m_2]=1+q_2.
\]
Moreover, $m_2$ belongs to every player's action-stage
information. The law of iterated expectations
therefore gives
\[
\mathbb E[E_i[X]\mid m_2]
=
\mathbb E[X\mid m_2].
\]
Taking expectations in \eqref{eq:action-focs}
conditional on $m_2$ yields
\[
\begin{aligned}
2\mathbb E[a_1\mid m_2]
&=1+q_2+b_1+\mathbb E[a_2\mid m_2],\\
3\mathbb E[a_2\mid m_2]
&=1+q_2+b_2
+\mathbb E[a_1\mid m_2]
+\mathbb E[a_3\mid m_2],\\
2\mathbb E[a_3\mid m_2]
&=1+q_2+b_3+\mathbb E[a_2\mid m_2].
\end{aligned}
\]
Substituting the first and third equations into the
second gives
\[
\begin{aligned}
3\mathbb E[a_2\mid m_2]
={}&1+q_2+b_2\\
&+\frac{1+q_2+b_1+\mathbb E[a_2\mid m_2]}2\\
&+\frac{1+q_2+b_3+\mathbb E[a_2\mid m_2]}2.
\end{aligned}
\]
Hence,
\[
2\mathbb E[a_2\mid m_2]
=
2+2q_2+b_2+\frac{b_1+b_3}{2},
\]
and therefore
\begin{equation}\label{eq:centre-common-mean}
\mathbb E[a_2\mid m_2]
=
1+q_2+\frac{b_1+2b_2+b_3}{4}.
\end{equation}
Substitution into the other two equations gives
\begin{equation}\label{eq:leaf-common-means}
\begin{aligned}
\mathbb E[a_1\mid m_2]
&=1+q_2+\frac{5b_1+2b_2+b_3}{8},\\
\mathbb E[a_3\mid m_2]
&=1+q_2+\frac{b_1+2b_2+5b_3}{8}.
\end{aligned}
\end{equation}
These are averages conditional only on $m_2$;
they are not the players' realized actions.

\medskip
\noindent
\textit{3. Deriving the leaves' actions.}

Taking expectations in player $1$'s best-response
equation conditional on $(m_1,m_2)$ gives
\begin{equation}\label{eq:leaf-conditional-equation}
2\mathbb E[a_1\mid m_1,m_2]
=
q_1+q_2+\frac12+b_1
+\mathbb E[a_2\mid m_1,m_2].
\end{equation}

Player $3$'s action is a function of her own type
and the messages she observes. She does not observe
$m_1$. Independence therefore implies
\[
\mathbb E[a_3\mid m_1,m_2]
=
\mathbb E[a_3\mid m_2]
=
1+q_2+\frac{b_1+2b_2+5b_3}{8}.
\]
Taking expectations in player $2$'s best-response
equation conditional on $(m_1,m_2)$ consequently gives
\begin{equation}\label{eq:centre-conditional-equation}
\begin{aligned}
3\mathbb E[a_2\mid m_1,m_2]
={}&q_1+q_2+\frac12+b_2
+\mathbb E[a_1\mid m_1,m_2]\\
&+1+q_2+\frac{b_1+2b_2+5b_3}{8}.
\end{aligned}
\end{equation}
The tower property is valid here because both
$m_1$ and $m_2$ belong to player $2$'s information.

Substituting \eqref{eq:leaf-conditional-equation}
into \eqref{eq:centre-conditional-equation} and
multiplying by two yields
\[
\begin{aligned}
6\mathbb E[a_2\mid m_1,m_2]
={}&2q_1+2q_2+1+2b_2\\
&+q_1+q_2+\frac12+b_1
+\mathbb E[a_2\mid m_1,m_2]\\
&+2+2q_2+\frac{b_1+2b_2+5b_3}{4}.
\end{aligned}
\]
Collecting terms,
\[
\begin{aligned}
5\mathbb E[a_2\mid m_1,m_2]
={}&3q_1+5q_2+\frac72\\
&+b_1+2b_2+\frac{b_1+2b_2+5b_3}{4}\\
={}&3q_1+5q_2+\frac72
+\frac54(b_1+2b_2+b_3).
\end{aligned}
\]
Thus,
\[
\mathbb E[a_2\mid m_1,m_2]
=
\frac35q_1+q_2+\frac7{10}
+\frac{b_1+2b_2+b_3}{4}.
\]

Player $2$ observes $m_1$, rather than $k_1$.
Conditional on $(m_1,m_2)$, player $1$'s additional
knowledge of $k_1$ does not change the distribution
of the variables determining player $2$'s action.
Hence,
\begin{equation}\label{eq:E1a2}
E_1[a_2]
=
\frac35q_1+q_2+\frac7{10}
+\frac{b_1+2b_2+b_3}{4}.
\end{equation}

Substitution into player $1$'s best response gives
\[
\begin{aligned}
2a_1
={}&k_1+q_2+\frac12+b_1
+\frac35q_1+q_2+\frac7{10}
+\frac{b_1+2b_2+b_3}{4}\\
={}&k_1+\frac35q_1+2q_2+\frac65
+\frac{5b_1+2b_2+b_3}{4}.
\end{aligned}
\]
Therefore,
\begin{equation}\label{eq:leaf1-action}
a_1
=
\frac12k_1+\frac3{10}q_1+q_2+\frac35
+\frac{5b_1+2b_2+b_3}{8}.
\end{equation}
By symmetry,
\begin{equation}\label{eq:leaf3-action}
a_3
=
\frac12k_3+\frac3{10}q_3+q_2+\frac35
+\frac{b_1+2b_2+5b_3}{8}.
\end{equation}

\medskip
\noindent
\textit{4. Deriving player $2$'s action.}

Player $2$ observes both leaves' messages, so
\[
E_2[k_1]=q_1,
\qquad
E_2[k_3]=q_3.
\]
Using \eqref{eq:leaf1-action},
\[
\begin{aligned}
E_2[a_1]
&=\frac12q_1+\frac3{10}q_1+q_2+\frac35
+\frac{5b_1+2b_2+b_3}{8}\\
&=\frac45q_1+q_2+\frac35
+\frac{5b_1+2b_2+b_3}{8}.
\end{aligned}
\]
Similarly,
\[
E_2[a_3]
=
\frac45q_3+q_2+\frac35
+\frac{b_1+2b_2+5b_3}{8}.
\]
Substituting into player $2$'s best response,
\[
\begin{aligned}
3a_2
={}&q_1+k_2+q_3+b_2\\
&+\frac45q_1+q_2+\frac35
+\frac{5b_1+2b_2+b_3}{8}\\
&+\frac45q_3+q_2+\frac35
+\frac{b_1+2b_2+5b_3}{8}\\
={}&k_2+2q_2+\frac95(q_1+q_3)+\frac65\\
&+b_2+\frac{6b_1+4b_2+6b_3}{8}\\
={}&k_2+2q_2+\frac95(q_1+q_3)+\frac65
+\frac34(b_1+2b_2+b_3).
\end{aligned}
\]
Dividing by three gives
\[
a_2
=
\frac13k_2+\frac23q_2
+\frac35(q_1+q_3)+\frac25
+\frac{b_1+2b_2+b_3}{4}.
\]

These calculations derive all three actions jointly.
Each player's conditional expected utility is strictly
concave in her own action. Furthermore, conditional
expectation is a contraction in $L^2$, and the
continuation best-response mapping has contraction
modulus at most $2/3$ under the maximum component
$L^2$ norm. The displayed actions therefore constitute
the unique square-integrable continuation equilibrium.

\medskip
\noindent
\textit{5. Player $1$'s cutoff equation.}

Fix $k_1=s$ and consider adjacent reports inducing
\[
q^+:=\frac{p_{1,r}+p_{1,r+1}}2,
\qquad
q^-:=\frac{p_{1,r-1}+p_{1,r}}2.
\]
Under these reports, player $2$'s actions are
\[
a_2^\pm
=
\frac13k_2+\frac23q_2
+\frac35(q^\pm+q_3)+\frac25
+\frac{b_1+2b_2+b_3}{4}.
\]
Player $1$ expects
\[
E_1[a_2^\pm]
=
\frac35q^\pm+q_2+\frac7{10}
+\frac{b_1+2b_2+b_3}{4},
\]
and optimally chooses
\[
a_1^\pm
=
\frac12s+\frac3{10}q^\pm+q_2+\frac35
+\frac{5b_1+2b_2+b_3}{8}.
\]
The relevant conditional variances are
\[
\operatorname{Var}_1(\theta)
=
\operatorname{Var}(k_2\mid m_2)+\frac1{12}
\]
and
\[
\operatorname{Var}_1(a_2^\pm)
=
\frac19\operatorname{Var}(k_2\mid m_2)
+\frac9{25}\operatorname{Var}(q_3).
\]
Consequently,
\[
\begin{aligned}
E_1[u_1^\pm]
={}&
-\left(a_1^\pm-s-q_2-\frac12-b_1\right)^2
-\left(a_1^\pm-E_1[a_2^\pm]\right)^2\\
&-\frac{10}{9}\operatorname{Var}(k_2\mid m_2)
-\frac1{12}
-\frac9{25}\operatorname{Var}(q_3).
\end{aligned}
\]
The variance terms do not depend on player $1$'s
chosen report. Substituting the actions gives
\[
\begin{aligned}
a_1^\pm-s-q_2-\frac12-b_1
&=-\frac12s+\frac3{10}q^\pm+\frac1{10}
-\frac{3b_1-2b_2-b_3}{8}\\
&=\frac12\left[
\frac35q^\pm-s+\frac15
-\frac{3b_1-2b_2-b_3}{4}
\right],
\end{aligned}
\]
and
\[
a_1^\pm-E_1[a_2^\pm]
=
-\frac12\left[
\frac35q^\pm-s+\frac15
-\frac{3b_1-2b_2-b_3}{4}
\right].
\]
The difference-of-squares identity therefore yields
\begin{equation}\label{eq:leaf-payoff-difference}
\begin{aligned}
E_1[u_1^+]-E_1[u_1^-]
={}&-\frac9{50}(q^+-q^-)
\Bigg[
q^++q^--\frac{10}{3}s+\frac23\\
&\hspace{3cm}
-\frac56(3b_1-2b_2-b_3)
\Bigg].
\end{aligned}
\end{equation}
This difference is independent of the incoming message
$m_2$, so it also gives the communication-stage
payoff comparison. At $s=p_{1,r}$, adjacent indifference implies
\[
\frac{p_{1,r-1}+2p_{1,r}+p_{1,r+1}}2
=
\frac{10}{3}p_{1,r}-\frac23
+\frac56(3b_1-2b_2-b_3).
\]
Multiplying by six,
\[
3p_{1,r-1}+6p_{1,r}+3p_{1,r+1}
=
20p_{1,r}-4+5(3b_1-2b_2-b_3).
\]
Rearranging gives \eqref{eq:leaf-recurrence}.

These adjacent indifference conditions also ensure optimality
against nonadjacent interval reports. Indeed, after optimizing
her continuation action, player $1$ ranks posterior means $q$
according to
\[
-\frac9{50}\left[
q-\left(\frac53s-\frac13
+\frac5{12}(3b_1-2b_2-b_3)\right)
\right]^2.
\]
The preferred posterior mean is strictly increasing in $s$.
For ordered interval means, the boundary between two adjacent
choices is their midpoint. Equation~\eqref{eq:leaf-recurrence}
places this boundary at the prescribed cutoff. Consequently,
every type in an interval weakly prefers its assigned report
to every other interval report. Cutoff types are indifferent
between adjacent reports and may be assigned to either.
Unused messages can be assigned the same beliefs and responses
as an existing interval message, so they introduce no additional
profitable deviations for player $1$. This establishes the
leaf's reporting incentives, conditional on the other players'
specified strategies; a full communication equilibrium also
requires their communication-stage optimality.

\medskip
\noindent
\textit{6. Player $1$'s first cutoff.}

The characteristic equation of
\eqref{eq:leaf-recurrence} is
\[
3z^2-14z+3=0.
\]
Its roots are $\lambda$ and $\lambda^{-1}$, where
\[
\lambda:=\frac{7+2\sqrt{10}}3>1.
\]
A constant particular solution is
\[
\frac{4-5(3b_1-2b_2-b_3)}8.
\]
Solving the recurrence subject to
$p_{1,0}=0$ and $p_{1,R_1}=1$ gives
\begin{equation}\label{eq:leaf-cutoffs}
\begin{aligned}
p_{1,r}
={}&
\frac{4-5(3b_1-2b_2-b_3)}8
\left[
1-
\frac{\lambda^{R_1-r}-\lambda^{-(R_1-r)}}
     {\lambda^{R_1}-\lambda^{-R_1}}
\right]\\
&+
\frac{4+5(3b_1-2b_2-b_3)}8
\frac{\lambda^r-\lambda^{-r}}
     {\lambda^{R_1}-\lambda^{-R_1}}.
\end{aligned}
\end{equation}
In particular,
\begin{equation}\label{eq:leaf-first-cutoff}
\begin{aligned}
p_{1,1}
={}&
\frac{4-5(3b_1-2b_2-b_3)}8
\left[
1-
\frac{\lambda^{R_1-1}-\lambda^{-(R_1-1)}}
     {\lambda^{R_1}-\lambda^{-R_1}}
\right]\\
&+
\frac{4+5(3b_1-2b_2-b_3)}8
\frac{\lambda-\lambda^{-1}}
     {\lambda^{R_1}-\lambda^{-R_1}}.
\end{aligned}
\end{equation}

\medskip
\noindent
\textit{7. Player $1$'s maximum number of intervals.}

Using $M_1>0$ and $\lambda$ defined in the statement,
the cutoff equation becomes
\[
3p_{1,r+1}-14p_{1,r}+3p_{1,r-1}=8M_1.
\]
Equivalently,
\[
p_{1,r+1}-p_{1,r}
=
p_{1,r}-p_{1,r-1}
+\frac83p_{1,r}+\frac83M_1.
\]
Under the maintained assumption, starting from
$p_{1,0}=0$ and $p_{1,1}>0$, this equation implies
inductively that all subsequent interval lengths
are positive. Thus, admissibility is equivalent
to $p_{1,1}>0$.

The solution satisfying $p_{1,0}=0$ and
$p_{1,R_1}=1$ is
\[
p_{1,r}
=
-M_1
+
\frac{
M_1\left(\lambda^{R_1-r}-\lambda^{-(R_1-r)}\right)
+(1+M_1)(\lambda^r-\lambda^{-r})
}{
\lambda^{R_1}-\lambda^{-R_1}
}.
\]
In particular,
\[
p_{1,1}
=
\frac{
(1+M_1)(\lambda-\lambda^{-1})
-
M_1\left[
\lambda^{R_1}-\lambda^{-R_1}
-\lambda^{R_1-1}+\lambda^{-(R_1-1)}
\right]
}{
\lambda^{R_1}-\lambda^{-R_1}
}.
\]
Since the denominator is positive, $p_{1,1}>0$
is equivalent to
\[
\begin{aligned}
(1+M_1)(\lambda-\lambda^{-1})
>{}&
M_1\left[
\lambda^{R_1}-\lambda^{-R_1}
-\lambda^{R_1-1}+\lambda^{-(R_1-1)}
\right].
\end{aligned}
\]
Using
\[
\lambda-\lambda^{-1}
=
\frac{(\lambda-1)(\lambda+1)}{\lambda}
\]
and
\[
\begin{aligned}
&\lambda^{R_1}-\lambda^{-R_1}
-\lambda^{R_1-1}+\lambda^{-(R_1-1)}\\
&\qquad
=(\lambda-1)
\left(\lambda^{R_1-1}+\lambda^{-R_1}\right),
\end{aligned}
\]
we obtain
\[
\lambda^{R_1}+\lambda^{1-R_1}
<
\left(1+\frac1{M_1}\right)(\lambda+1).
\]

Set $x:=\lambda^{R_1}$. Multiplying the preceding
inequality by $x>0$ gives
\[
x^2
-
\left(1+\frac1{M_1}\right)(\lambda+1)x
+\lambda
<0.
\]
The larger root is
\[
x^*
=
\frac{
\left(1+\frac1{M_1}\right)(\lambda+1)
+
\sqrt{
\left(1+\frac1{M_1}\right)^2(\lambda+1)^2
-4\lambda
}
}{2}.
\]
The quadratic is positive at $x=0$ and negative
at $x=1$, because
\[
1-\left(1+\frac1{M_1}\right)(\lambda+1)+\lambda
=
-\frac{\lambda+1}{M_1}<0.
\]
Hence its smaller root lies between $0$ and $1$.
Since $x=\lambda^{R_1}\geq\lambda>1$, admissibility
reduces to
\[
\lambda^{R_1}<x^*.
\]
Taking logarithms,
\[
R_1<\frac{\log x^*}{\log\lambda}.
\]
Therefore,
\begin{equation}\label{eq:leaf-maximum}
R_1^{\max}
=
\left\lceil
\frac{\log x^*}{\log\lambda}
\right\rceil-1.
\end{equation}
The ceiling minus one accounts for the strict
positivity of the first interval. As $M_1\to0^+$,
\[
x^*\longrightarrow\infty,
\qquad
R_1^{\max}\longrightarrow\infty.
\]
In terms of the original biases, this limit corresponds to
\[
3b_1-2b_2-b_3\longrightarrow\frac45
\quad\text{from above}.
\]

\medskip
\noindent
\textit{8. Player $2$'s conditional comparison.}

Fix $k_2=s$ and adjacent outgoing reports inducing
\[
q^+:=\frac{p_{2,r}+p_{2,r+1}}2,
\qquad
q^-:=\frac{p_{2,r-1}+p_{2,r}}2.
\]
For the conditional comparison, write
\[
P:=\mathbb E[k_1\mid m_1],
\qquad
Q:=\mathbb E[k_3\mid m_3].
\]
These are the posterior means induced by the leaves'
reports. Hold $P,Q$ fixed when comparing the two
outgoing reports. The leaves' actual actions are
\[
a_1^\pm
=
\frac12k_1+\frac3{10}P+q^\pm+\frac35
+\frac{5b_1+2b_2+b_3}{8},
\]
\[
a_3^\pm
=
\frac12k_3+\frac3{10}Q+q^\pm+\frac35
+\frac{b_1+2b_2+5b_3}{8}.
\]
Player $2$ expects
\[
E_2[a_1^\pm]
=
\frac45P+q^\pm+\frac35
+\frac{5b_1+2b_2+b_3}{8},
\]
\[
E_2[a_3^\pm]
=
\frac45Q+q^\pm+\frac35
+\frac{b_1+2b_2+5b_3}{8}.
\]
Her expected state is $s+P+Q$, so her optimal
continuation action is
\[
a_2^\pm
=
\frac13s+\frac23q^\pm+\frac35(P+Q)+\frac25
+\frac{b_1+2b_2+b_3}{4}.
\]
The conditional expected payoff is
\[
\begin{aligned}
E_2[u_2^\pm]
={}&-\left(a_2^\pm-s-P-Q-b_2\right)^2\\
&-\left(a_2^\pm-E_2[a_1^\pm]\right)^2
-\left(a_2^\pm-E_2[a_3^\pm]\right)^2\\
&-\frac54\left[
\operatorname{Var}(k_1\mid m_1)
+\operatorname{Var}(k_3\mid m_3)
\right].
\end{aligned}
\]
The three deviations are
\[
\begin{aligned}
a_2^\pm-s-P-Q-b_2
={}&-\frac23s+\frac23q^\pm
-\frac25(P+Q)+\frac25\\
&+\frac{b_1-2b_2+b_3}{4},
\end{aligned}
\]
\[
\begin{aligned}
a_2^\pm-E_2[a_1^\pm]
={}&\frac13s-\frac13q^\pm
-\frac15P+\frac35Q-\frac15\\
&+\frac{-3b_1+2b_2+b_3}{8},
\end{aligned}
\]
and
\[
\begin{aligned}
a_2^\pm-E_2[a_3^\pm]
={}&\frac13s-\frac13q^\pm
+\frac35P-\frac15Q-\frac15\\
&+\frac{b_1+2b_2-3b_3}{8}.
\end{aligned}
\]
The variance terms cancel in the payoff comparison.
Applying the difference-of-squares identity gives
\[
\begin{aligned}
&\frac{E_2[u_2^+]-E_2[u_2^-]}{q^+-q^-}\\
&\quad=
-\frac23\left[
-\frac43s+\frac23(q^++q^-)
-\frac45(P+Q)+\frac45
+\frac{b_1-2b_2+b_3}{2}
\right]\\
&\qquad+
\frac13\left[
\frac43s-\frac23(q^++q^-)
+\frac45(P+Q)-\frac45
+\frac{-b_1+2b_2-b_3}{2}
\right].
\end{aligned}
\]
Collecting terms,
\begin{equation}\label{eq:centre-conditional-comparison}
\begin{aligned}
E_2[u_2^+]-E_2[u_2^-]
={}&-\frac23(q^+-q^-)
\Bigg[
q^++q^--2s-\frac65(P+Q-1)\\
&\hspace{3cm}
-\frac34(2b_2-b_1-b_3)
\Bigg].
\end{aligned}
\end{equation}
Holding $P,Q$ fixed, conditional indifference gives
\[
q^++q^-
=
2s+\frac65(P+Q-1)
+\frac34(2b_2-b_1-b_3).
\]
At a conditional cutoff $s=p_{2,r}$, this would imply
\begin{equation}\label{eq:conditional-centre-recurrence}
p_{2,r+1}-2p_{2,r}+p_{2,r-1}
=
\frac{12}{5}(P+Q-1)
+\frac32(2b_2-b_1-b_3).
\end{equation}

\medskip
\noindent
\textit{9. Player $2$'s first cutoff and maximum number
of intervals, holding $P,Q$ fixed.}

Consider
\[
p_{2,r+1}-2p_{2,r}+p_{2,r-1}
=
\frac{12}{5}(P+Q-1)
+\frac32(2b_2-b_1-b_3),
\]
with
\[
p_{2,0}=0,
\qquad
p_{2,R_2}=1.
\]
Successive interval lengths satisfy
\[
\begin{aligned}
p_{2,r+1}-p_{2,r}
={}&p_{2,r}-p_{2,r-1}\\
&+\frac{12}{5}(P+Q-1)
+\frac32(2b_2-b_1-b_3).
\end{aligned}
\]
Therefore,
\[
p_{2,r}-p_{2,r-1}
=
p_{2,1}
+(r-1)
\left[
\frac{12}{5}(P+Q-1)
+\frac32(2b_2-b_1-b_3)
\right].
\]
Summing from $1$ to $r$ gives
\[
p_{2,r}
=
r p_{2,1}
+
\left[
\frac65(P+Q-1)
+\frac34(2b_2-b_1-b_3)
\right]r(r-1).
\]
Using $p_{2,R_2}=1$,
\[
1
=
R_2p_{2,1}
+
\left[
\frac65(P+Q-1)
+\frac34(2b_2-b_1-b_3)
\right]R_2(R_2-1).
\]
Hence, the first cutoff is
\begin{equation}\label{eq:conditional-centre-first-cutoff}
p_{2,1}
=
\frac1{R_2}
-
\left[
\frac65(P+Q-1)
+\frac34(2b_2-b_1-b_3)
\right](R_2-1).
\end{equation}
Substituting back gives
\begin{equation}\label{eq:conditional-centre-cutoffs}
p_{2,r}
=
\frac r{R_2}
+
\left[
\frac65(P+Q-1)
+\frac34(2b_2-b_1-b_3)
\right]r(r-R_2).
\end{equation}
Thus, the interval lengths are
\[
p_{2,r}-p_{2,r-1}
=
\frac1{R_2}
+
\left[
\frac65(P+Q-1)
+\frac34(2b_2-b_1-b_3)
\right](2r-R_2-1).
\]
Under the maintained positive-right-hand-side
assumption, these lengths increase with $r$.
Consequently, all intervals have positive length
if and only if $p_{2,1}>0$. By
\eqref{eq:conditional-centre-first-cutoff},
\[
\frac1{R_2}
-
\left[
\frac65(P+Q-1)
+\frac34(2b_2-b_1-b_3)
\right](R_2-1)>0.
\]
Equivalently,
\[
R_2(R_2-1)
<
\frac{20}{
24(P+Q-1)+15(2b_2-b_1-b_3)
}.
\]
Solving this quadratic inequality gives
\[
R_2
<
\frac12+
\frac12\sqrt{
1+
\frac{80}{
24(P+Q-1)+15(2b_2-b_1-b_3)
}
}.
\]
Therefore, the maximum number of positive-length
intervals for the conditional cutoff system is
\begin{equation}\label{eq:conditional-centre-maximum}
R_2^{\max}(P,Q)
=
\left\lceil
\frac12+
\frac12\sqrt{
1+
\frac{80}{
24(P+Q-1)+15(2b_2-b_1-b_3)
}
}
\right\rceil-1.
\end{equation}
The strict inequality explains the ceiling minus one.
This completes the conditional calculation. Because messages
are simultaneous, player $2$ does not observe $P,Q$ when
choosing her report. Her actual communication-stage incentive
constraint must average the conditional payoff comparison over
these incoming reports. We do not replace that constraint by
conditional indifference at each $P,Q$.
\end{proof}


\begin{thebibliography}{99}

\bibitem{Ambrus2013}
\textbf{Ambrus, A.}, \textbf{Azevedo, E. M.}, \& \textbf{Kamada, Y.} (2013). Hierarchical cheap talk. \textit{Theoretical Economics, 8}(1), 233-261.

\bibitem{Ballester2006}
\textbf{Ballester, C.}, \textbf{Calvó‐Armengol, A.}, \& \textbf{Zenou, Y.} (2006). Who's who in networks. Wanted: The key player. \textit{Econometrica, 74}(5), 1403-1417.

\bibitem{Bonacich1987}
\textbf{Bonacich, P.} (1987). Power and centrality: A family of measures. \textit{American Journal of Sociology, 92}(5), 1170-1182.

\bibitem{Chen2008}
\textbf{Chen, Y.}, \textbf{Kartik, N.}, \& \textbf{Sobel, J.} (2008). Selecting cheap‐talk equilibria. \textit{Econometrica, 76}(1), 117-136.

\bibitem{Crawford1982}
\textbf{Crawford, V. P.}, \& \textbf{Sobel, J.} (1982). Strategic information transmission. \textit{Econometrica: Journal of the Econometric Society}, 1431-1451.

\bibitem{Dasaratha2020}
\textbf{Dasaratha, K.} (2020). Distributions of centrality on networks. \textit{Games and Economic Behavior, 122}, 1-27.

\bibitem{deMartiZenou2015}
\textbf{de Mart\'i, J., \& Zenou, Y.} (2015). Network games with incomplete information. \textit{Journal of Mathematical Economics, 61}, 221--240.

\bibitem{Galeotti2013}
\textbf{Galeotti, A.}, \textbf{Ghiglino, C.}, \& \textbf{Squintani, F.} (2013). Strategic information transmission networks. \textit{Journal of Economic Theory, 148}(5), 1751-1769.

\bibitem{Ghiglino2010}
\textbf{Ghiglino, C.}, \& \textbf{Goyal, S.} (2010). Keeping up with the neighbors: social interaction in a market economy. \textit{Journal of the European Economic Association, 8}(1), 90-119.

\bibitem{Hagenbach2010}
\textbf{Hagenbach, J.}, \& \textbf{Koessler, F.} (2010). Strategic communication networks. \textit{The Review of Economic Studies, 77}(3), 1072-1099.

\bibitem{Immorlica2017}
\textbf{Immorlica, N.}, \textbf{Kranton, R.}, \textbf{Manea, M.}, \& \textbf{Stoddard, G.} (2017). Social status in networks. \textit{American Economic Journal: Microeconomics, 9}(1), 1-30.

\bibitem{Sadler2020}
\textbf{Sadler, E.} (2020). Ordinal Centrality. Available at SSRN. 

\bibitem{UshchevZenou2020}
\textbf{Ushchev, P., \& Zenou, Y.} (2020). Social norms in networks. \textit{Journal of Economic Theory, 185}, 104969.

\end{thebibliography}
\end{document}